\documentclass[twocolumn,amsmath,amssymb,prl,superscriptaddress]{revtex4-1}                    %Reference without titles
\usepackage{titlesec}
\usepackage[colorlinks,bookmarks=false,citecolor=blue,linkcolor=red,urlcolor=blue]{hyperref}
\usepackage{epsfig}
\usepackage{color}
\usepackage{bm}
\usepackage{subfigure}
\usepackage{graphicx}    % Include figure files
\usepackage{dcolumn}     % Align table columns on decimal point
\usepackage{lipsum}      % to move around 2-column tables !

\input{insbox}

\newcommand{\be}{\begin{equation}}
\newcommand{\ee}{\end{equation}}

\definecolor{drkgr}{rgb}{0.05,0.6,0.2}

\begin{document}

\title{NaRuO$_2$: Kitaev-Heisenberg exchange in triangular-lattice setting}

\author{Pritam Bhattacharyya}
\affiliation{Institute for Theoretical Solid State Physics, Leibniz IFW Dresden, Helmholtzstra{\ss}e~20, 01069 Dresden, Germany}

\author{Nikolay A. Bogdanov}
\affiliation{Max Planck Institute for Solid State Research, Heisenbergstra{\ss}e~1, 70569 Stuttgart, Germany}

\author{Satoshi Nishimoto}
\affiliation{Institute for Theoretical Solid State Physics, Leibniz IFW Dresden, Helmholtzstra{\ss}e~20, 01069 Dresden, Germany}
\affiliation{Department of Physics, Technical University Dresden, 01069 Dresden, Germany}

\author{Stephen D.~Wilson}
\affiliation{Materials Department, University of California, Santa Barbara, California 93106-5050, USA}

\author{Liviu Hozoi}
\affiliation{Institute for Theoretical Solid State Physics, Leibniz IFW Dresden, Helmholtzstra{\ss}e~20, 01069 Dresden, Germany}

\begin{abstract}
\noindent
Kitaev exchange, a new paradigm in quantum magnetism research, occurs for 90$^{\circ}$ metal-ligand-metal
links, $t_{2g}^5$ transition ions, and sizable spin-orbit coupling.
It is being studied in honeycomb compounds but also on triangular lattices.
While for the former it is known by now that the Kitaev intersite couplings are ferromagnetic, for the
latter the situation is unclear.
Here we pin down the exchange mechanisms and determine the effective coupling constants in the $t_{2g}^5$
triangular-lattice material NaRuO$_2$, recently found to host a quantum spin liquid ground state. 
We show that, compared to honeycomb compounds, the characteristic triangular-lattice cation surroundings
dramatically affect exchange paths and effective coupling parameters, changing the Kitaev interactions to
antiferromagnetic.
The quantum chemical analysis and subsequent effective spin model computations provide perspective onto
the nature of the experimentally observed quantum spin liquid --- it seemingly implies finite longer-range
exchange, and the atypical proximity to ferromagnetic order is related to sizable ferromagnetic Heisenberg
nearest-neighbor couplings.
\end{abstract}

%\date\today
\maketitle

\begin{figure}[b]
\includegraphics[width=0.95\columnwidth]{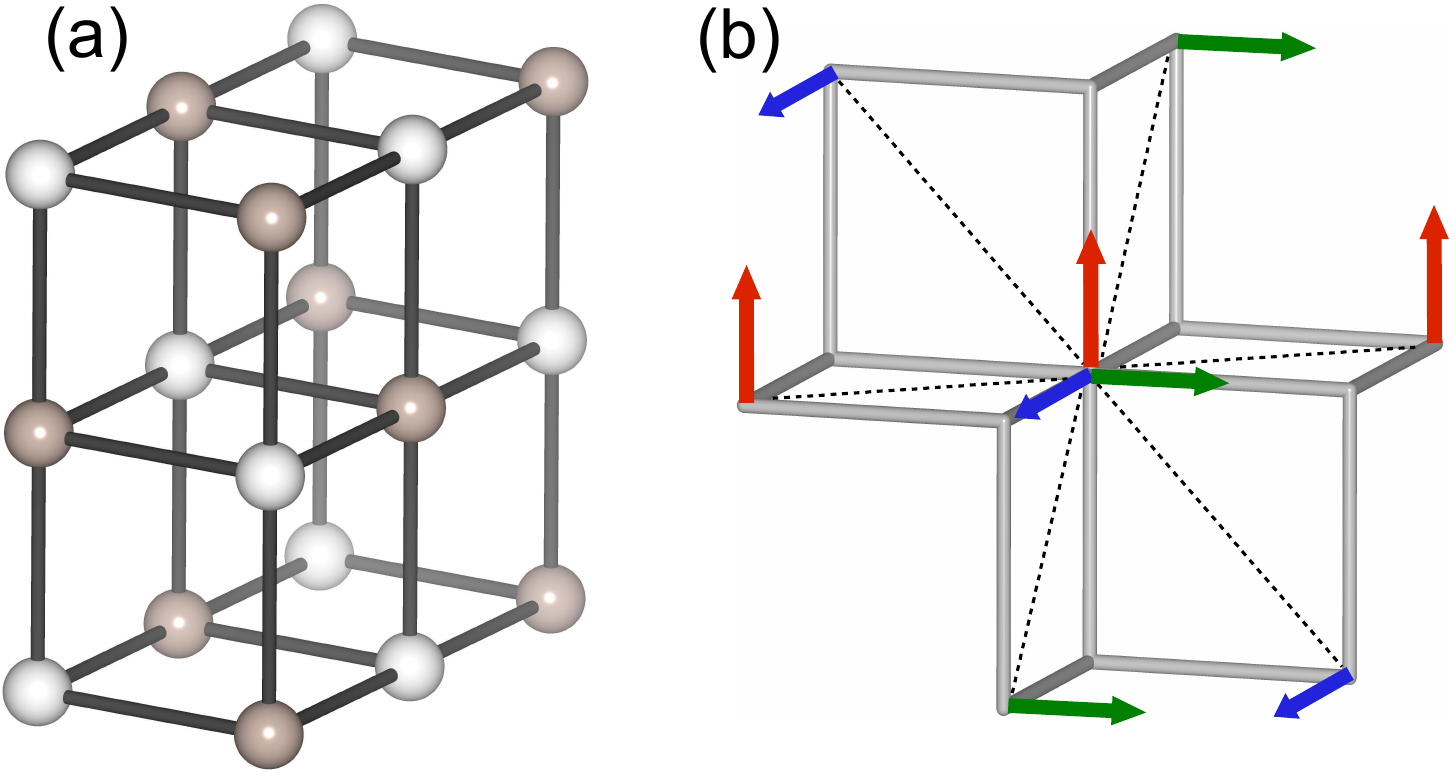}
\caption{
(a) Sketch of the NaCl-type lattice, with M-L bonds along $x$, $y$, or $z$.
If two different cation varieties (alkaline, $A$, and TM, $B$, ions) form successive layers
normal to the [111] axis, a rhombohedral $AB$L$_2$ structure is realized, with a triangular
network of edge-sharing octahedra in each layer.
(b) On each $B_2$L$_2$ plaquette (ions not drawn), the Kitaev interaction couples only (pseudo)spin
components perpendicular to the respective plaquette.
All three components are shown only for the central magnetic site.
}
\label{kitaev_spins}
\end{figure}

%% \noindent
%% {\it Introduction.\,}
Exchange through 90$^{\circ}$ metal-ligand-metal bonds represents one of the limiting cases in
(super)exchange theory \cite{KANAMORI1959,Anderson_1959}.
In the simplest situation of half-filled $d$-metal orbitals, this geometry is associated with
Heisenberg ferromagnetism.
Away from half-filling and for degenerate orbitals, however, very intricate physics may arise, as
pointed out by Jackeli and Khaliullin for $t_{2g}^5$ magnetic centers with sizable spin-orbit
coupling: highly anisotropic effective interactions involving only (pseudo)spin components normal
to the M$_2$L$_2$ square plaquette formed by two transition-metal (TM) ions and the two bridging
ligands \cite{Ir213_KH_jackeli_09,Takagi2019}.
In crystals of NaCl type and derivative rhombohedral structures which imply three different possible
orientations of the M$_2$L$_2$ plaquettes (see \ref{kitaev_spins}), this translates into directional dependence
of the nearest-neighbor exchange: 
on differently oriented adjacent M$_2$L$_2$ units --- i.e., normal to either $x$, $y$, or $z$ ---
different components of the magnetic moments interact, either
$\tilde{S}_{i}^{x}$ and $\tilde{S}_{j}^{x}$ (on $x$-type, normal to the $x$-axis plaquettes),
$\tilde{S}_{i}^{y}$ and $\tilde{S}_{k}^{y}$ ($y$ type), or
$\tilde{S}_{i}^{z}$ and $\tilde{S}_{l}^{z}$ ($z$ type).
The (super)exchange model of Jackeli and Khaliullin \cite{Ir213_KH_jackeli_09} thus formalizes
Kitaev's effective Hamiltonian of bond-dependent anisotropic magnetic couplings \cite{Kitaev2006}
proposed initially more like a heuristic device.
It launched a whole new subfield in the research area of quantum magnetic materials \cite{D1TC02070F,
Cava_2021}, that of 5$d^5$ and 4$d^5$ honeycomb magnets \cite{Takagi2019}, with subsequent extension
to 3$d^7$ $t_{2g}^5e_g^2$ hexagonal networks of edge-sharing ML$_6$ octahedra.

%%%%%%%%%%%%%%%%%
%%% FIGURE 1 %%%%
%%%%%%%%%%%%%%%%%

Here we explore the nature of nearest-neighbor couplings in a 4$d^5$ {\it triangular}-lattice magnet,
NaRuO$_2$, and evidence the presence of sizable bond-dependent Kitaev interactions.
Interestingly, those are antiferromagnetic, different from the case of honeycomb 4$d^5$ and 5$d^5$
magnetic lattices.
Preliminary numerical tests highlight the important role of cation species around the ligands
mediating (super)exchange: the sign change is related to electrostatics having to do with adjacent 
TM ions that in the honeycomb setting are missing.
Sizable antiferromagnetic off-diagonal intersite couplings are also predicted, along with somewhat
larger ferromagnetic Heisenberg exchange.
The latter seems to be consistent qualitatively with features seen in experiment: incipient
ferromagnetism within a quantum disordered ground state \cite{ru112_Ortiz_2022,ru112_Ortiz_2022_2}.
By quantum chemical computations at different levels of approximation, we additionally show that
anisotropic Coulomb exchange defines a major interaction scale: it represents $\sim$45\% of the
kinetic exchange contribution in the case of the diagonal Kitaev coupling $K$ and is 4 times larger
than $d$-$d$ kinetic exchange for the off-diagonal $\Gamma'$.
Coulomb exchange being ignored so far in Kitaev (super)exchange models, these results provide 
perspective onto what reliable quantitative predictions would imply: not only controlled approximations
to deal with intersite virtual hopping \cite{KANAMORI1959,Anderson_1959,Ir213_KH_jackeli_09} but also
exact Coulomb exchange.

%%%%%%%%%%%%%%%%%
%%% FIGURE 2 %%%%
%%%%%%%%%%%%%%%%%
\begin{figure}[b]
\includegraphics[width=0.90\columnwidth]{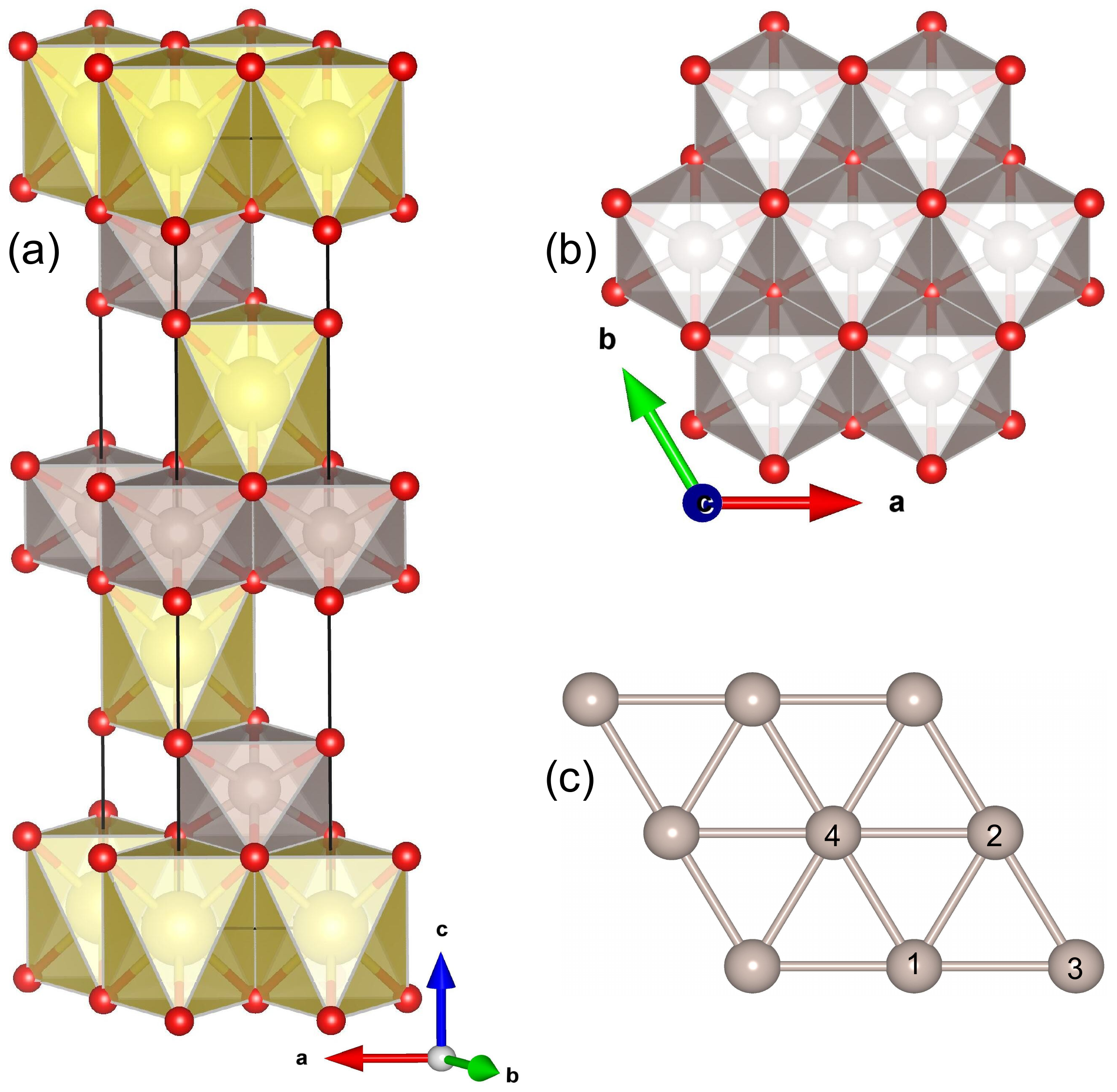}
\caption{
Delafossite layered structure (a) and the triangular network of edge-sharing octahedra of NaRuO$_2$
(b);
grey, red, and yellow spheres represent Ru, O, and Na ions, respectively.
(c) To understand how the nearby surroundings affect magnetic couplings, Ru neighbors in the median
plane (sites 3, 4) of a given Ru-Ru magnetic link (sites 1, 2) were removed in one set of quantum
chemical computations.
}
\label{ABL2_structure}
\end{figure}

\

\noindent
{\bf Results}\\[0.12cm]
{\it On-site multiplet structure and intersite couplings in NaRuO$_2$.\,}
NaRuO$_2$ is rather unique, a realization of $j_{\mathrm{eff}}\!\approx\!1/2$ $t_{2g}^5$ moments on a
triangular network of edge-sharing RuO$_6$ octahedra (see \ref{ABL2_structure}).
The basic Ru$^{3+}$ 4$d^5$ multiplet structure in the specific delafossite crystalline setting of NaRuO$_2$
is illustrated in \ref{d5}, as obtained by quantum chemical \cite{olsen_bible} complete-active-space
self-consistent-field (CASSCF) \cite{olsen_bible,MCSCF_Molpro} and post-CASSCF multireference
configuration-interaction (MRCI) \cite{olsen_bible,MRCI_Molpro} embedded-cluster calculations (see
Methods section for computational details and Supplementary Information for basis sets).
To separately visualize the effects of crystal-field (CF) splittings and spin-orbit coupling (SOC),
both scalar relativistic (CASSCF, MRCI) and MRCI+SOC \cite{SOC_Molpro} results are listed. 
The $t_{2g}$-$e_g$ splitting is larger than in e.\,g. RuCl$_3$ \cite{Yadav2016}, due to larger ligand
effective charges and shorter TM-ligand bonds in NaRuO$_2$.
It is seen that the MRCI correlation treatment brings significant corrections to some of the CASSCF
relative energies, in particular for the $^6\!A_{1g}$ term.
Also important is the SOC (see the last two columns in \ref{d5}).
However, the effects of trigonal fields are visible too: those remove the degeneracy of the $^2T_{2g}$
CF states ($p$-orbital-like $l_{\mathrm{eff}}\!=\!1/2$ \cite{book_abragam_bleaney} and $s\!=\!1/2$
quantum numbers in ideal cubic environment) and speak for significant deviations from textbook \cite{
book_abragam_bleaney,Ir213_KH_jackeli_09} $j_{\mathrm{eff}}\!=\!1/2$ spin-orbit moments.

Even with sizable trigonal fields, existing estimates of several meV for the Kitaev interactions in
honeycomb-lattice 4$d$ $t_{2g}^5$ compounds such as RuCl$_3$ (see e.\,g.~\cite{RuCl3_ML} and discussion
in \cite{rucl3_janssen_17}) and Li$_2$RhO$_3$ \cite{rh213_vmk_15} provide strong motivation for detailed
{\it ab initio} investigation of the effective couplings in NaRuO$_2$.
To this end, further quantum chemical computations were performed for a block of two adjacent
edge-sharing RuO$_6$ octahedra, following the procedure described in Refs.~\cite{Ir214_niko_15,
Yadav2016} (see also Methods section).
% The accuracy of the magnetic interactions, both isotropic and anisotropic, derived through this
% methodology was confirmed by comparison to estimates extacted from experimental data, see
% e.\,g.~Ref.~\cite{Ir214_niko_15}.
%
Mapping the quantum chemical data onto the relevant effective spin model for $C_{2h}$ symmetry of
the [Ru$_2$O$_{10}$] magnetic unit,

\begin{equation}
\label{eqn:Hamil}
\mathcal{H}_{ij}^{(\gamma)} = J{\bf{\tilde{S}}}_i\cdot {\bf{\tilde{S}}}_j + \\
                              K\tilde{S}_i^{\gamma}\tilde{S}_j^{\gamma} + \\
                              \sum_{\alpha\neq\beta} \Gamma_{\alpha\beta} \\
                              (\tilde{S}_i^{\alpha}\tilde{S}_j^{\beta} +  \\
                              \tilde{S}_i^{\beta}\tilde{S}_j^{\alpha}) ,
\end{equation}
by spin-orbit MRCI, we arrive to Kitaev, Heisenberg, and off-diagonal $\Gamma$ coupling constants
$K\!=\!2.0$, $J\!=\!-5.2$, $\Gamma\!\equiv\!\Gamma_{xy}\!=\!3.6$,
$\Gamma'\!\equiv\!\Gamma_{yz}\!=\!\Gamma_{zx}\!=\!1.1$ (meV).
Interestingly, both $K$ and $\Gamma$ are antiferromagnetic, with $\Gamma\!>\!K$, but the largest
interaction parameter is the Heisenberg $J$.
The latter being ferromagnetic, it brings us away from the antiferromagnetic ground states (zigzag
antiferromagnetic order, in particular) experimentally found in $j\!\approx\!1/2$ honeycomb systems,
as depicted for instance in the phase diagrams computed for $J$-$K$-$\Gamma$ triangular-lattice
models in Ref.~\cite{PD}. 

%%%%%%%%%%%%%
%% TABLE 1 %%
%%%%%%%%%%%%%
\begin{table}[t]
\caption{
Ru$^{3+}$ $4d^5$ multiplet structure, with all five $4d$ orbitals active in CASSCF.
Each MRCI+SOC value refers to a Kramers doublet (KD);
just the lowest and highest KDs are shown for each group of $t_{2g}^4e_g^1$ spin-orbit states.
Only the CF terms listed in the table entered the spin-orbit calculations.
For simplicity, notations corresponding to $O_h$ symmetry are used.
}
\begin{tabular}{llll}
\hline
\hline\\[-0.20cm]
Ru$^{3+}$ $4d^5$               &CASSCF     &MRCI   &MRCI\\
splittings (eV)               &           &       &+SOC\\
\hline
\\[-0.30cm]
$^2T_{2g}$ ($t_{2g}^5$)        &0          &0      &0    \\
                               &0.11       &0.12   &0.21\\
                               &0.11       &0.12   &0.26\\[0.10cm]
$^4T_{1g}$ ($t_{2g}^4e_g^1$)   &1.42       &1.58   &1.61\\
                               &1.45       &1.62   &$|$ \\
                               &1.45       &1.62   &1.74\\[0.10cm]
$^6\!A_{1g}$ ($t_{2g}^3e_g^2$) &1.42       &1.91   &2.08\,($\times 3$)\\[0.10cm]
$^4T_{2g}$ ($t_{2g}^4e_g^1$)   &2.16       &2.28   &2.37\\
                               &2.16       &2.28   &$|$ \\
                               &2.20       &2.30   &2.42\\
\hline
\hline
\end{tabular}
\label{d5}
\end{table}

%%%%%%%%%%%%%
%% TABLE 2 %%
%%%%%%%%%%%%%
\begin{table}[!b!]
\caption{
Nearest-neighbor effective magnetic couplings (MRCI+SOC) for experimentally determined crystal
structures of NaRuO$_2$\,/\,RuCl$_3$ and for adjusted configurations with modified median plane
(MMP) cation distribution (see main text and \ref{ABL2_structure}(c) for details).
}
\begin{tabular}{l r r r r}
\hline
\hline \\[-0.20cm]

MRCI+SOC (meV)            &$K$   &$J$    &$\Gamma$  &$\Gamma'$\\

\hline \\[-0.20cm]

NaRuO$_2$                 &2.0   &--5.2  &3.6       &1.1      \\
NaRuO$_2$, MMP            &--19  &12     &1.2       &--1.1    \\[0.15cm]
RuCl$_3$ \cite{Yadav2016} &--5.6 &1.2    &1.2       &--0.7    \\
RuCl$_3$, MMP             &1.9   &--12   &1.1       &--0.1    \\
\hline
\hline
\end{tabular}
\label{table_nn_cations}
\end{table}

According to Ref.~\cite{PD}, quantum spin liquid (QSL) ground states can only be realized for relatively
large, antiferromagnetic $\Gamma$.
Our result for the strength of $\Gamma$ represents the largest {\it ab initio} quantum chemical estimate
so far in real material setting.
Yet, a nearest-neighbor $J$ that is even larger in magnitude and ferromagnetic (see the values listed
above and at the top of \ref{table_nn_cations}) implies that only longer-range Heisenberg
couplings and/or higher-order interactions such as ring exchange \cite{ring_exchange_motrunich_05}
can pull the system to the QSL regime evidenced experimentally \cite{ru112_Ortiz_2022}. 

Signatures of a nearby ferromagnetic state in NaRuO$_2$ appear via the enhanced Van Vleck term in the
magnetic susceptibility \cite{ru112_Ortiz_2022}, which approaches values observed in nearly magnetic
metals like Pd. 
Furthermore, despite possessing a charge gap and insulating ground state, clear quasiparticle excitations
generate a Sommerfeld-like term in the low temperature heat capacity \cite{ru112_Ortiz_2022}.
This suggests the presence of strong spin fluctuations associated with a nearby magnetic state.
These seemingly gapless spin fluctuations are robust, and are directly resolved in neutron scattering as
diffuse continuum-like modes about the nuclear zone center and near \textbf{Q}=0 \cite{ru112_Ortiz_2022}.
Initial attempts at scaling S(Q,E) suggest the proximity of a nearby ferromagnetic quantum critical
point, and small, nonmagnetic chemical perturbations to the network of Ru$^{3+}$ ions induce a glass-like
freezing of the magnetic moments \cite{ru112_Ortiz_2022_2}.

%%%%%%%%%%%%%%%%%
%%% FIGURE 3 %%%%
%%%%%%%%%%%%%%%%%
\begin{figure*}[!t!]
\includegraphics[width=1.0\textwidth]{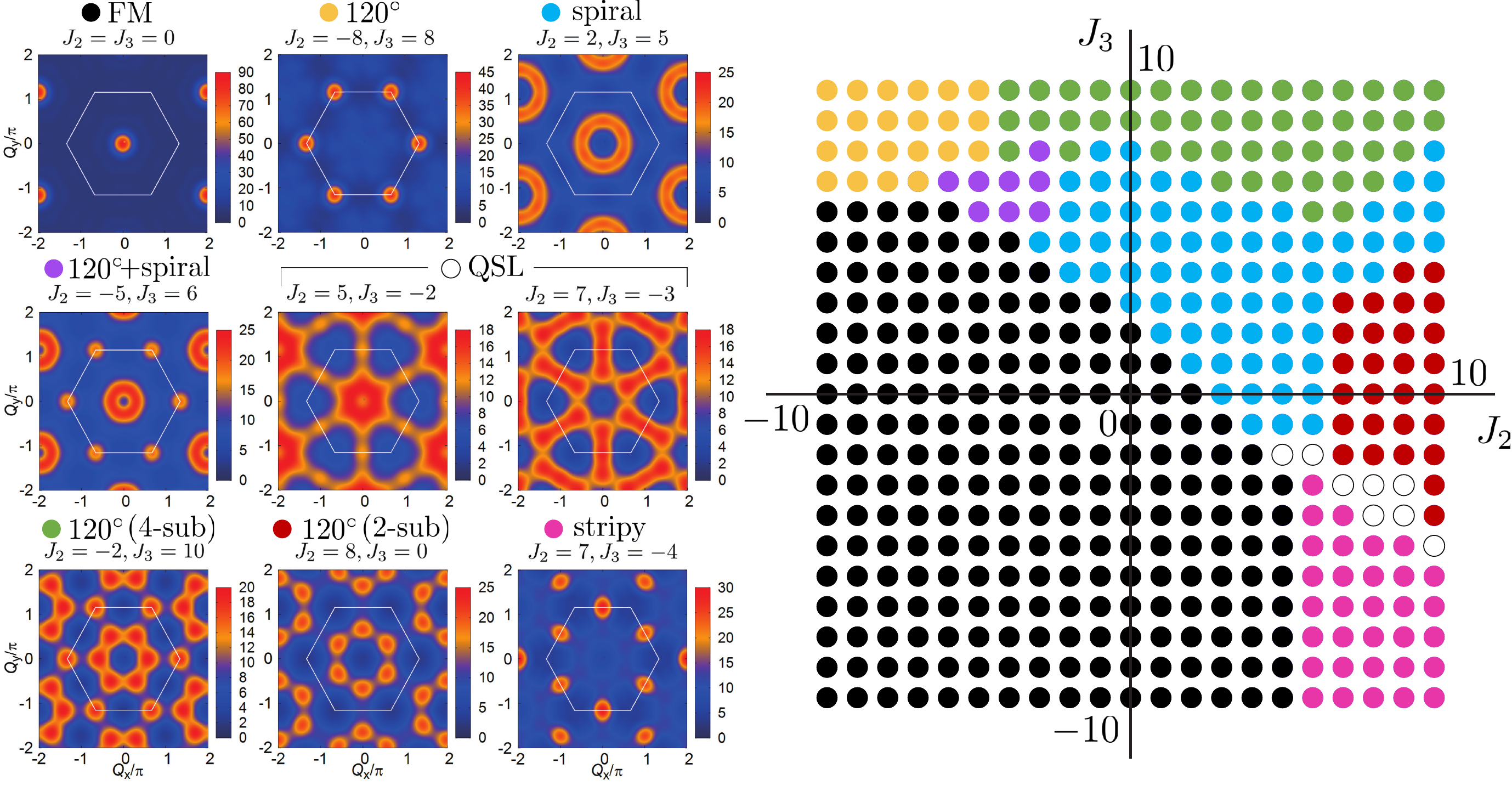}
\caption{
Ground-state phase diagram obtained by DMRG computations for the $J$-$K$-$\Gamma$-$\Gamma'$-$J_2$-$J_3$
model (right) and typical static spin structure factors for each phase (left).
Using the quantum chemical analysis as a guide (see data in  \ref{table_correlations} and related
discussion), $J$, $K$, $\Gamma$, and $\Gamma'$ were set to to --5.5, 4, 4, and 1.5 meV, respectively.
}
\label{fig_dmrg}
\end{figure*}

{\it What makes $K$ antiferromagnetic.\,}
Finding an antiferromagnetic Kitaev coupling in NaRuO$_2$, opposite to the ferromagnetic $K$ generally
found in honeycomb Kitaev-Heisenberg magnets (see \cite{rh213_vmk_15,Yadav2016,Yadav_CS} for quantum
chemical results and e.\,g.~\cite{RuCl3_ML,J3_winter_16} plus discussion in \cite{rucl3_janssen_17}
for altenative approaches), is intriguing.
To shed light on this aspect, we performed the following numerical experiment:
in a new set of quantum chemical computations, the two magnetic-plane cations in the immediate
neighborhood of the bridging ligands, forming a line perpendicular to the link of nearest-neighbor
Ru sites (see \ref{ABL2_structure}(c)), were removed and their charge redistributed within the point
charge array modeling the extended solid state surroundings.
Remarkably, without those nearby positive ions, three of the effective magnetic couplings change
sign (see  \ref{table_nn_cations}).
This suggests that strong orbital polarization effects at the ligand sites, induced by extra nearby
positive charge on the triangular lattice (+3 effective nearby charges in NaRuO$_2$ versus +1 in
honeycomb `213' compounds such as Na$_2$IrO$_3$ and Li$_2$RhO$_3$ or 0 in RuCl$_3$) have dramatic
consequences on hopping matrix elements and superexchange processes.
Similar additional tests for RuCl$_3$, where +3 ions were placed in the centers of two edge-sharing
Ru$_6$ hexagonal rings, confirm the trend: the extra positive charge in the neighborhood of the
Ru$_2$O$_2$ `magnetic' plaquette (i.\,e., [Ru$_2$O$_{10}$] cluster of edge-sharing octahedra) invert 
the sign of the Kitaev coupling constant, from ferromagnetic in actual RuCl$_3$ \cite{Yadav2016} to
antiferromagnetic in the presence of positive charge in the centers of the two hexagons sharing the
magnetic Ru-Ru link (see lowest lines in  \ref{table_nn_cations}).
Strong polarization effects of this type were earlier pointed out in the case of the Kitaev-Heisenberg
honeycomb iridate H$_3$LiIr$_2$O$_6$ \cite{Yadav_CS}, produced in that case by cations residing
between the magnetic layers.

%%%%%%%%%%%%%
%% TABLE 3 %%
%%%%%%%%%%%%%
\begin{table}[!t!]
\caption{
Nearest-neighbor magnetic couplings (meV) at different levels of theory:
SC (only the $t_{2g}^5$--$t_{2g}^5$ configuration considered), CASSCF (also $t_{2g}^4$--$t_{2g}^6$
states included), and MRCI (single and double excitations out of the Ru 4$d$ $t_{2g}$ and bridging-O
2$p$ levels on top of CASSCF).
}
\begin{tabular}{l r r r r}
\hline
\hline \\[-0.20cm]

Magnetic couplings   &$K$   &$J$    &$\Gamma$  &$\Gamma'$\\

\hline \\[-0.20cm]

SC+SOC               &--1.0 &--0.7  &0.4       &0.8      \\
CASSCF+SOC           &1.3   &--5.2  &2.2       &1.0      \\
MRCI+SOC             &2.0   &--5.2  &3.6       &1.1      \\
\hline
\hline
\end{tabular}
\label{table_correlations}
\end{table}

{\it Ru$_2$O$_2$-unit correlations. Extended magnetic lattice.\,}
Insights into the interplay between spin-orbit couplings and electron correlation effects on a 
plaquette of two nearest-neighbor Ru ions and two bridging ligands can be obtained by analysing
computational data obtained at lower levels of approximation, i.\,e., spin-orbit calculations
including only the $t_{2g}^5$--$t_{2g}^5$ electron configuration and CASSCF+SOC calculations that
also account for excited-state $t_{2g}^4$--$t_{2g}^6$ configurations since all possible occupations
are considered within the six-orbital (three $t_{2g}$ orbitals per Ru site) active space in the
latter case.

As illustrated in \ref{table_correlations}, anisotropic effective intersite interactions 
comparable in size with the isotropic Heisenberg $J$ are already found at the single-configuration
(SC) level, i.\,e., when accounting for just $t_{2g}^5$--$t_{2g}^5$ Coulomb exchange (sometimes 
referred to as direct exchange \cite{Anderson_1959}).
Large anisotropic Coulomb exchange as found in the SC calculation represents physics not addressed 
so far in the literature --- Kitaev magnetism is presently exclusively explained through TM-TM kinetic
exchange and TM-L-TM superexchange.

CASSCF, through inclusion of intersite $t_{2g}$$\rightarrow$\,$t_{2g}$ excitations (i.\,e., TM-TM
kinetic exchange), brings sizable corrections to $K$, $J$, and $\Gamma$ in particular.
By considering next, in the spin-orbit MRCI treatment, all possible single and double excitations 
out of the Ru $t_{2g}$ and bridging-ligand orbitals, the most significant post-CASSCF corrections
are found to occur for $K$ and $\Gamma$.
This suggests that more sophisticated calculations, e.\,g., MRCI+SOC based on larger active spaces
in the prior CASSCF, would bring significant additional corrections to $K$ and $\Gamma$ mostly, i.\,e., 
would only enhance antiferromagnetic fluctuations on the extended lattice.
The effects of sizable antiferromagnetic $\Gamma'$ (last column in \ref{table_correlations}),
longer-range Heisenberg couplings, and ring exchange \cite{ring_exchange_motrunich_05} are so far
unknown in triangular-lattice setting.

% Clarifying such aspects requires rather extensive additional investigations, combining complementary
% computational approaches.
% But to obtain first impressions on
To obtain first impressions on the role of longer-range Heisenberg interactions, second-neighbor
($J_2$) and third-neighbor ($J_3$), in particular, we performed density-matrix renormalization group
(DMRG) calculations \cite{DMRG-1,DMRG-2} for a $J$-$K$-$\Gamma$-$\Gamma'$-$J_2$-$J_3$ model on
fragments of 19, 27, and 37 sites of the triangular lattice.
In order to prevent artifacts, e.\,g., artificial stabilization (or destabilization) of particular
magnetic states, we applied open boundary conditions.
The validity of this material model is discussed in the Supplementary Information.
Setting $J$, $K$, $\Gamma$, and $\Gamma'$ to --5.5, 4, 4, and 1.5 meV, respectively, we obtain a
very rich phase diagram, see \ref{fig_dmrg}.
Notably, a QSL phase is found for antiferromagnetic $J_2$ values and ferromagnetic $J_3$.
The corresponding structure factor has no Bragg peaks but exhibits a characteristic pattern.
It indicates that the QSL emerges from the competition of adjacent ordered states.
The most remarkable spot is the region where the QSL neighbors ferromagnetic order and a spiral phase,
namely, around $J_2\!=\!5$, $J_3\!=\!-2$ --- 
as shown in \ref{fig_dmrg}, at $J_2\!=\!5$, $J_3\!=\!-2$ the structure factor implies a broad 
spectral feature for momenta near $|$\textbf{Q}$|$=0, seemingly consistent with experimental observations
\cite{ru112_Ortiz_2022}.
Interestingly, ferromagnetic and long-period `spiral' domains in the presence of defects would also
yield a broad spectral feature in the vicinity of $|$\textbf{Q}$|$=0.
The rather sharp increase seen at low fields in the magnetization curve \cite{ru112_Ortiz_2022} would
be reproduced in either case.

\

\

\noindent
{\bf Conclusions}\\[0.12cm]
While the way nearby cations are structurally arranged can strongly affect on-site electronic-structure 
features such as subshell level splittings \cite{Os227_niko_12,Ir214_niko_15,yb13_ziba_2019}, single-ion
anisotropies \cite{Os227_niko_12}, and $g$ factors \cite{Ir214_niko_15,yb13_ziba_2019}, detailed
{\it ab initio} investigations of the effect on intersite magnetic couplings are less numerous.
Here we show that, compared to honeycomb compounds, the characteristic triangular-lattice cation
surroundings dramatically affect superexchange paths and effective coupling parameters.
% , through polarization of the bridging-ligand valence electronic charge.
In particular, with respect to honeycomb $j\!\approx\!1/2$ systems, the Kitaev interaction constant
changes its sign in triangular-lattice NaRuO$_2$.
By providing insight into the signs and strengths of the nearest-neighbor magnetic interactions in
this material, our work sets the frame within which the role of longer-range effective spin couplings
should be addressed.
Interestingly, while giving rise to antiferromagnetic order on honeycomb Kitaev-Heisenberg lattices,
the latter appear to be decisive in realizing quantum disorder \cite{ru112_Ortiz_2022} in NaRuO$_2$.
Last but not least, we establish the role of anisotropic Coulomb exchange, a mechanism not addressed
so far in Kitaev-Heisenberg magnetism. 

\

\noindent
{\bf Methods}\\[0.12cm]
{\it Ru-site multiplet structure.\,}
All quantum chemical computations were carried out using the {\sc molpro} suite of programs \cite{
Molpro}.
Crystallographic data as reported by Ortiz {\it et al.}~\cite{ru112_Ortiz_2022} were utilized.
For each type of embedded cluster, the crystalline environment was modeled as a large array of point
charges which reproduces the crystalline Madelung field within the cluster volume; 
we employed the {\sc ewald} program \cite{Klintenberg_et_al,Derenzo_et_al} to generate the point-charge
embeddings.

To clarify the Ru-site multiplet structure, a cluster consisting of one `central' RuO$_6$ octahedron,
the six in-plane adjacent octahedra, and 12 nearby Na cations was designed.
The quantum chemical study was initiated as complete active space self-consistent field (CASSCF)
calculations \cite{olsen_bible,MCSCF_Molpro}, with an active orbital space containing the five 4$d$
orbitals of the central Ru ion.
Post-CASSCF correlation computations were performed at the level of multireference configuration-interaction
(MRCI) with single and double excitations \cite{olsen_bible,MRCI_Molpro} out of the Ru 4$d$ and O 2$p$
orbitals of the central RuO$_6$ octahedron.
Spin-orbit couplings (SOCs) were accounted for following the procedure described in Ref.~\cite{SOC_Molpro}. 

{\it Intersite exchange in NaRuO$_2$.\,}
Clusters with two edge-sharing RuO$_6$ octahedra in the central region were considered in order to
to derive the inter-site effective magnetic couplings.
The eight in-plane octahedra directly linked to the two-octahedra central unit and 22 nearby Na ions
were also explicitly included in the quantum chemical computations but using much more compact basis
sets.

CASSCF computations were carried out with six (Ru $t_{2g}$) valence orbitals and ten electrons as
active (abbreviated as (10e,6o) active space);
the $t_{2g}$ orbitals of the adjacent TM ions were part of the inactive orbital space.
In the subsequent MRCI correlation treatment, single and double excitations out of the central-unit
Ru $t_{2g}$ and bridging-O 2$p$ levels were considered.
We used the Pipek-Mezey methodology \cite{PM_method} to obtain localized central-unit orbitals.

The CASSCF optimization was performed for the lowest nine singlet and lowest nine triplet states
associated with the (10e,6o) setting.
Those were the states for which SOCs were further accounted for \cite{SOC_Molpro}, either at 
single-configuration (SC), CASSCF, or MRCI level, which finally yields a number of 36 spin-orbit
states.
The SC label in Table III in the main text stands for a CASCI in which intersite excitations 
are not considered.
This is also referred to as occupation-restricted multiple active space (ORMAS) scheme \cite{
ormas_ivanic_03}.

Only one type of Ru-Ru links is present in NaRuO$_2$.
The unit of two nearest-neighbor octahedra displays $C_{2h}$ point-group symmetry, which implies a 
generalized bilinear effective spin Hamiltonian of the following form for a pair of adjacent
1/2-pseudospins ${\bf{\tilde{S}}}_i$ and ${\bf{\tilde{S}}}_j$\,:

\begin{equation}
\label{eqn:Hamil}
\mathcal{H}_{ij}^{(\gamma)} = J{\bf{\tilde{S}}}_i\cdot {\bf{\tilde{S}}}_j + \\
                              K\tilde{S}_i^{\gamma}\tilde{S}_j^{\gamma} + \\
                              \sum_{\alpha\neq\beta} \Gamma_{\alpha\beta} \\
                              (\tilde{S}_i^{\alpha}\tilde{S}_j^{\beta} +  \\
                              \tilde{S}_i^{\beta}\tilde{S}_j^{\alpha}).
\end{equation}
The $\Gamma_{\alpha\beta}$ coefficients refer to the off-diagonal components of the 3$\times$3
symmetric-anisotropy exchange matrix;
$\alpha,\beta,\gamma\!\in\!\{x,y,z\}$.
An antisymmetric Dzyaloshinskii-Moriya coupling is not allowed, given the inversion center.
% A local Kitaev reference frame is used here, such that for each Ru--Ru link the $z$ axis is
% perpendicular to the `magnetic' Ru$_2$O$_2$ plaquette.

The lowest four spin-orbit eigenstates from the {\sc molpro} output (eigenvalues lower by $\sim$200
meV or more than the eigenvalues of higher-lying excited states, as illustrated for example in Table\;I)
were mapped onto the eigenvectors of the effective spin Hamiltonian (\ref{eqn:Hamil}),
following the procedure described in Refs.~\cite{Bogdanov_et_al,Yadav2016}\,:
those four expectation values and the matrix elements of the Zeeman Hamiltonian in the basis of the
four lowest-energy spin-orbit eigenvectors are put in direct correspondence with the respective
eigenvalues and matrix elements of (\ref{eqn:Hamil}).
Having two of the states in the same irreducible representation of the $C_{2h}$ point group \cite{
rh213_vmk_15}, such one-to-one mapping translates into two possible sets of effective magnetic
couplings.
The relevant array is chosen as the one whose $g$ factors fit the $g$ factors obtained 
for a single RuO$_6$ $t_{2g}^5$ octahedron. 

We used the standard coordinate frame usually employed in the literature, different from the rotated
frame employed in earlier quantum chemical studies \cite{Yadav2016,rh213_vmk_15,Yadav_CS} that
affects the sign of $\Gamma$.
This is the reason the sign of $\Gamma$ for RuCl$_3$ in Table II in the main text is different from
the sign in Ref.~\cite{Yadav2016} (see also footnote [48] in  Ref.~\cite{rucl3_janssen_17}).

\

{\bf Data Availability.\,}
The data that support the findings of the current study are available from the corresponding author 
upon reasonable request.

\

{\bf Acknowledgments.\,}
P.\,B.~and L.\,H.~acknowledge financial support from the German Research Foundation (Deutsche 
Forschungsgemeinschaft, DFG), Project ID 441216021, and technical assistance from U.\,Nitzsche.
S.\,N.~acknowledges financial support via project A05 of the Collaborative Research Center 
1143 of the DFG (Project ID 247310070).
S.\,D.\,W.~acknowledges support via DOE, Office of Science, Basic Energy Sciences under Award
DE-SC0017752.
P.\,B.~thanks M.\,S.\,Eldeeb for discussions.
We also acknowledge instructive discussions with I.\,Rousochatzakis, U.\,K.\,R{\"o}{\ss}ler,
and G.\,Khaliullin.

 \

{\bf Author Contributions.\,}
P.\,B.~performed the {\it ab initio} quantum chemical computations, with assistance from 
L.\,H., and N.\,A.\,B.
S.\,N.~carried out the model-Hamiltonian calculations for the extended lattice and subsequent analysis.
S.\,D.\,W.~and L.\,H.~initiated this study.
P.\,B., S.\,N., S.\,D.\,W. and L.\,H.~wrote the paper, with contributions from all other coauthors.

{\bf Competing Interests.\,}
The authors declare no competing interests.

\bibliography{refs_apr19}

\begin{thebibliography}{33}%
\makeatletter
\providecommand \@ifxundefined [1]{%
 \@ifx{#1\undefined}
}%
\providecommand \@ifnum [1]{%
 \ifnum #1\expandafter \@firstoftwo
 \else \expandafter \@secondoftwo
 \fi
}%
\providecommand \@ifx [1]{%
 \ifx #1\expandafter \@firstoftwo
 \else \expandafter \@secondoftwo
 \fi
}%
\providecommand \natexlab [1]{#1}%
\providecommand \enquote  [1]{``#1''}%
\providecommand \bibnamefont  [1]{#1}%
\providecommand \bibfnamefont [1]{#1}%
\providecommand \citenamefont [1]{#1}%
\providecommand \href@noop [0]{\@secondoftwo}%
\providecommand \href [0]{\begingroup \@sanitize@url \@href}%
\providecommand \@href[1]{\@@startlink{#1}\@@href}%
\providecommand \@@href[1]{\endgroup#1\@@endlink}%
\providecommand \@sanitize@url [0]{\catcode `\\12\catcode `\$12\catcode
  `\&12\catcode `\#12\catcode `\^12\catcode `\_12\catcode `\%12\relax}%
\providecommand \@@startlink[1]{}%
\providecommand \@@endlink[0]{}%
\providecommand \url  [0]{\begingroup\@sanitize@url \@url }%
\providecommand \@url [1]{\endgroup\@href {#1}{\urlprefix }}%
\providecommand \urlprefix  [0]{URL }%
\providecommand \Eprint [0]{\href }%
\providecommand \doibase [0]{http://dx.doi.org/}%
\providecommand \selectlanguage [0]{\@gobble}%
\providecommand \bibinfo  [0]{\@secondoftwo}%
\providecommand \bibfield  [0]{\@secondoftwo}%
\providecommand \translation [1]{[#1]}%
\providecommand \BibitemOpen [0]{}%
\providecommand \bibitemStop [0]{}%
\providecommand \bibitemNoStop [0]{.\EOS\space}%
\providecommand \EOS [0]{\spacefactor3000\relax}%
\providecommand \BibitemShut  [1]{\csname bibitem#1\endcsname}%
\let\auto@bib@innerbib\@empty
%</preamble>
\bibitem [{\citenamefont {Kanamori}(1959)}]{KANAMORI1959}%
  \BibitemOpen
  \bibfield  {author} {\bibinfo {author} {\bibfnamefont {J.}~\bibnamefont
  {Kanamori}},\ }\href {\doibase https://doi.org/10.1016/0022-3697(59)90061-7}
  {\bibfield  {journal} {\bibinfo  {journal} {J. Phys. Chem. Solids}\ }\textbf
  {\bibinfo {volume} {10}},\ \bibinfo {pages} {87} (\bibinfo {year}
  {1959})}\BibitemShut {NoStop}%
\bibitem [{\citenamefont {Anderson}(1959)}]{Anderson_1959}%
  \BibitemOpen
  \bibfield  {author} {\bibinfo {author} {\bibfnamefont {P.~W.}\ \bibnamefont
  {Anderson}},\ }\href {\doibase 10.1103/PhysRev.115.2} {\bibfield  {journal}
  {\bibinfo  {journal} {Phys. Rev.}\ }\textbf {\bibinfo {volume} {115}},\
  \bibinfo {pages} {2} (\bibinfo {year} {1959})}\BibitemShut {NoStop}%
\bibitem [{\citenamefont {Jackeli}\ and\ \citenamefont
  {Khaliullin}(2009)}]{Ir213_KH_jackeli_09}%
  \BibitemOpen
  \bibfield  {author} {\bibinfo {author} {\bibfnamefont {G.}~\bibnamefont
  {Jackeli}}\ and\ \bibinfo {author} {\bibfnamefont {G.}~\bibnamefont
  {Khaliullin}},\ }\href {\doibase 10.1103/PhysRevLett.102.017205} {\bibfield
  {journal} {\bibinfo  {journal} {Phys. Rev. Lett.}\ }\textbf {\bibinfo
  {volume} {102}},\ \bibinfo {pages} {017205} (\bibinfo {year}
  {2009})}\BibitemShut {NoStop}%
\bibitem [{\citenamefont {Takagi}\ \emph {et~al.}(2019)\citenamefont {Takagi},
  \citenamefont {Takayama}, \citenamefont {Jackeli}, \citenamefont
  {Khaliullin},\ and\ \citenamefont {Nagler}}]{Takagi2019}%
  \BibitemOpen
  \bibfield  {author} {\bibinfo {author} {\bibfnamefont {H.}~\bibnamefont
  {Takagi}}, \bibinfo {author} {\bibfnamefont {T.}~\bibnamefont {Takayama}},
  \bibinfo {author} {\bibfnamefont {G.}~\bibnamefont {Jackeli}}, \bibinfo
  {author} {\bibfnamefont {G.}~\bibnamefont {Khaliullin}}, \ and\ \bibinfo
  {author} {\bibfnamefont {S.~E.}\ \bibnamefont {Nagler}},\ }\href {\doibase
  10.1038/s42254-019-0038-2} {\bibfield  {journal} {\bibinfo  {journal} {Nat.
  Rev. Phys.}\ }\textbf {\bibinfo {volume} {1}},\ \bibinfo {pages} {264}
  (\bibinfo {year} {2019})}\BibitemShut {NoStop}%
\bibitem [{\citenamefont {Kitaev}(2006)}]{Kitaev2006}%
  \BibitemOpen
  \bibfield  {author} {\bibinfo {author} {\bibfnamefont {A.}~\bibnamefont
  {Kitaev}},\ }\href {\doibase 10.1016/j.aop.2005.10.005} {\bibfield  {journal}
  {\bibinfo  {journal} {Ann. Phys.}\ }\textbf {\bibinfo {volume} {321}},\
  \bibinfo {pages} {2} (\bibinfo {year} {2006})}\BibitemShut {NoStop}%
\bibitem [{\citenamefont {Browne}\ \emph {et~al.}(2021)\citenamefont {Browne},
  \citenamefont {Krajewska},\ and\ \citenamefont {Gibbs}}]{D1TC02070F}%
  \BibitemOpen
  \bibfield  {author} {\bibinfo {author} {\bibfnamefont {A.~J.}\ \bibnamefont
  {Browne}}, \bibinfo {author} {\bibfnamefont {A.}~\bibnamefont {Krajewska}}, \
  and\ \bibinfo {author} {\bibfnamefont {A.~S.}\ \bibnamefont {Gibbs}},\ }\href
  {\doibase 10.1039/D1TC02070F} {\bibfield  {journal} {\bibinfo  {journal} {J.
  Mater. Chem. C}\ }\textbf {\bibinfo {volume} {9}},\ \bibinfo {pages} {11640}
  (\bibinfo {year} {2021})}\BibitemShut {NoStop}%
\bibitem [{\citenamefont {Cava}\ \emph {et~al.}(2021)\citenamefont {Cava},
  \citenamefont {de~Leon},\ and\ \citenamefont {Xie}}]{Cava_2021}%
  \BibitemOpen
  \bibfield  {author} {\bibinfo {author} {\bibfnamefont {R.}~\bibnamefont
  {Cava}}, \bibinfo {author} {\bibfnamefont {N.}~\bibnamefont {de~Leon}}, \
  and\ \bibinfo {author} {\bibfnamefont {W.}~\bibnamefont {Xie}},\ }\href
  {\doibase 10.1021/acs.chemrev.0c01322} {\bibfield  {journal} {\bibinfo
  {journal} {Chem. Rev.}\ }\textbf {\bibinfo {volume} {121}},\ \bibinfo {pages}
  {2777} (\bibinfo {year} {2021})}\BibitemShut {NoStop}%
\bibitem [{\citenamefont {Ortiz}\ \emph {et~al.}(shed)\citenamefont {Ortiz},
  \citenamefont {Sarte}, \citenamefont {Avidor}, \citenamefont {Hay},
  \citenamefont {Kenney}, \citenamefont {Kolisneikov}, \citenamefont
  {Pajerowski}, \citenamefont {Aczel}, \citenamefont {Taddei}, \citenamefont
  {Brown}, \citenamefont {Wang}, \citenamefont {Graf}, \citenamefont
  {Seshadri}, \citenamefont {Balents},\ and\ \citenamefont
  {Wilson}}]{ru112_Ortiz_2022}%
  \BibitemOpen
  \bibfield  {author} {\bibinfo {author} {\bibfnamefont {B.~R.}\ \bibnamefont
  {Ortiz}}, \bibinfo {author} {\bibfnamefont {P.~M.}\ \bibnamefont {Sarte}},
  \bibinfo {author} {\bibfnamefont {A.~H.}\ \bibnamefont {Avidor}}, \bibinfo
  {author} {\bibfnamefont {A.}~\bibnamefont {Hay}}, \bibinfo {author}
  {\bibfnamefont {E.}~\bibnamefont {Kenney}}, \bibinfo {author} {\bibfnamefont
  {A.~I.}\ \bibnamefont {Kolisneikov}}, \bibinfo {author} {\bibfnamefont
  {D.~M.}\ \bibnamefont {Pajerowski}}, \bibinfo {author} {\bibfnamefont
  {A.~A.}\ \bibnamefont {Aczel}}, \bibinfo {author} {\bibfnamefont {K.~M.}\
  \bibnamefont {Taddei}}, \bibinfo {author} {\bibfnamefont {C.}~\bibnamefont
  {Brown}}, \bibinfo {author} {\bibfnamefont {C.}~\bibnamefont {Wang}},
  \bibinfo {author} {\bibfnamefont {M.~J.}\ \bibnamefont {Graf}}, \bibinfo
  {author} {\bibfnamefont {R.}~\bibnamefont {Seshadri}}, \bibinfo {author}
  {\bibfnamefont {L.}~\bibnamefont {Balents}}, \ and\ \bibinfo {author}
  {\bibfnamefont {S.~D.}\ \bibnamefont {Wilson}},\ }\href
  {https://www.researchsquare.com/article/rs-1551865/v1} {\bibfield  {journal}
  {\bibinfo  {journal} {Research Square rs-1551865}\ } (\bibinfo {year}
  {unpublished})}\BibitemShut {NoStop}%
\bibitem [{\citenamefont {Ortiz}\ \emph {et~al.}(2022)\citenamefont {Ortiz},
  \citenamefont {Sarte}, \citenamefont {Avidor},\ and\ \citenamefont
  {Wilson}}]{ru112_Ortiz_2022_2}%
  \BibitemOpen
  \bibfield  {author} {\bibinfo {author} {\bibfnamefont {B.~R.}\ \bibnamefont
  {Ortiz}}, \bibinfo {author} {\bibfnamefont {P.~M.}\ \bibnamefont {Sarte}},
  \bibinfo {author} {\bibfnamefont {A.~H.}\ \bibnamefont {Avidor}}, \ and\
  \bibinfo {author} {\bibfnamefont {S.~D.}\ \bibnamefont {Wilson}},\ }\href
  {\doibase https://doi.org/10.1103/PhysRevMaterials.6.104413} {\bibfield
  {journal} {\bibinfo  {journal} {Phys. Rev. Materials}\ }\textbf {\bibinfo
  {volume} {6}},\ \bibinfo {pages} {104413} (\bibinfo {year}
  {2022})}\BibitemShut {NoStop}%
\bibitem [{\citenamefont {Helgaker}\ \emph {et~al.}(2000)\citenamefont
  {Helgaker}, \citenamefont {J{\o}rgensen},\ and\ \citenamefont
  {Olsen}}]{olsen_bible}%
  \BibitemOpen
  \bibfield  {author} {\bibinfo {author} {\bibfnamefont {T.}~\bibnamefont
  {Helgaker}}, \bibinfo {author} {\bibfnamefont {P.}~\bibnamefont
  {J{\o}rgensen}}, \ and\ \bibinfo {author} {\bibfnamefont {J.}~\bibnamefont
  {Olsen}},\ }\href@noop {} {\emph {\bibinfo {title} {Molecular Electronic
  Structure Theory}}}\ (\bibinfo  {publisher} {John Wiley \& Sons},\ \bibinfo
  {address} {Chichester},\ \bibinfo {year} {2000})\BibitemShut {NoStop}%
\bibitem [{\citenamefont {Kreplin}\ \emph {et~al.}(2020)\citenamefont
  {Kreplin}, \citenamefont {Knowles},\ and\ \citenamefont
  {Werner}}]{MCSCF_Molpro}%
  \BibitemOpen
  \bibfield  {author} {\bibinfo {author} {\bibfnamefont {D.~A.}\ \bibnamefont
  {Kreplin}}, \bibinfo {author} {\bibfnamefont {P.~J.}\ \bibnamefont
  {Knowles}}, \ and\ \bibinfo {author} {\bibfnamefont {H.-J.}\ \bibnamefont
  {Werner}},\ }\href {\doibase 10.1063/1.5142241} {\bibfield  {journal}
  {\bibinfo  {journal} {J. Chem. Phys.}\ }\textbf {\bibinfo {volume} {152}},\
  \bibinfo {pages} {074102} (\bibinfo {year} {2020})}\BibitemShut {NoStop}%
\bibitem [{\citenamefont {Knowles}\ and\ \citenamefont
  {Werner}(1992)}]{MRCI_Molpro}%
  \BibitemOpen
  \bibfield  {author} {\bibinfo {author} {\bibfnamefont {P.~J.}\ \bibnamefont
  {Knowles}}\ and\ \bibinfo {author} {\bibfnamefont {H.-J.}\ \bibnamefont
  {Werner}},\ }\href {\doibase 10.1007/BF01117405} {\bibfield  {journal}
  {\bibinfo  {journal} {Theor. Chim. Acta}\ }\textbf {\bibinfo {volume} {84}},\
  \bibinfo {pages} {95} (\bibinfo {year} {1992})}\BibitemShut {NoStop}%
\bibitem [{\citenamefont {Berning}\ \emph {et~al.}(2000)\citenamefont
  {Berning}, \citenamefont {Schweizer}, \citenamefont {Werner}, \citenamefont
  {Knowles},\ and\ \citenamefont {Palmieri}}]{SOC_Molpro}%
  \BibitemOpen
  \bibfield  {author} {\bibinfo {author} {\bibfnamefont {A.}~\bibnamefont
  {Berning}}, \bibinfo {author} {\bibfnamefont {M.}~\bibnamefont {Schweizer}},
  \bibinfo {author} {\bibfnamefont {H.-J.}\ \bibnamefont {Werner}}, \bibinfo
  {author} {\bibfnamefont {P.~J.}\ \bibnamefont {Knowles}}, \ and\ \bibinfo
  {author} {\bibfnamefont {P.}~\bibnamefont {Palmieri}},\ }\href {\doibase
  10.1080/00268970009483386} {\bibfield  {journal} {\bibinfo  {journal} {Mol.
  Phys.}\ }\textbf {\bibinfo {volume} {98}},\ \bibinfo {pages} {1823} (\bibinfo
  {year} {2000})}\BibitemShut {NoStop}%
\bibitem [{\citenamefont {Yadav}\ \emph {et~al.}(2016)\citenamefont {Yadav},
  \citenamefont {Bogdanov}, \citenamefont {Katukuri}, \citenamefont
  {Nishimoto}, \citenamefont {van~den Brink},\ and\ \citenamefont
  {Hozoi}}]{Yadav2016}%
  \BibitemOpen
  \bibfield  {author} {\bibinfo {author} {\bibfnamefont {R.}~\bibnamefont
  {Yadav}}, \bibinfo {author} {\bibfnamefont {N.~A.}\ \bibnamefont {Bogdanov}},
  \bibinfo {author} {\bibfnamefont {V.~M.}\ \bibnamefont {Katukuri}}, \bibinfo
  {author} {\bibfnamefont {S.}~\bibnamefont {Nishimoto}}, \bibinfo {author}
  {\bibfnamefont {J.}~\bibnamefont {van~den Brink}}, \ and\ \bibinfo {author}
  {\bibfnamefont {L.}~\bibnamefont {Hozoi}},\ }\href {\doibase
  10.1038/srep37925} {\bibfield  {journal} {\bibinfo  {journal} {Sci. Rep.}\
  }\textbf {\bibinfo {volume} {6}},\ \bibinfo {pages} {37925} (\bibinfo {year}
  {2016})}\BibitemShut {NoStop}%
\bibitem [{\citenamefont {Abragam}\ and\ \citenamefont
  {Bleaney}(1970)}]{book_abragam_bleaney}%
  \BibitemOpen
  \bibfield  {author} {\bibinfo {author} {\bibfnamefont {A.}~\bibnamefont
  {Abragam}}\ and\ \bibinfo {author} {\bibfnamefont {B.}~\bibnamefont
  {Bleaney}},\ }\href@noop {} {\emph {\bibinfo {title} {Electron Paramagnetic
  Resonance of Transition Ions}}}\ (\bibinfo  {publisher} {Clarendon Press,
  Oxford},\ \bibinfo {year} {1970})\BibitemShut {NoStop}%
\bibitem [{\citenamefont {Samarakoon}\ \emph {et~al.}(2022)\citenamefont
  {Samarakoon}, \citenamefont {Laurell}, \citenamefont {Balz}, \citenamefont
  {Banerjee}, \citenamefont {Lampen-Kelley}, \citenamefont {Mandrus},
  \citenamefont {Nagler}, \citenamefont {Okamoto},\ and\ \citenamefont
  {Tennant}}]{RuCl3_ML}%
  \BibitemOpen
  \bibfield  {author} {\bibinfo {author} {\bibfnamefont {A.~M.}\ \bibnamefont
  {Samarakoon}}, \bibinfo {author} {\bibfnamefont {P.}~\bibnamefont {Laurell}},
  \bibinfo {author} {\bibfnamefont {C.}~\bibnamefont {Balz}}, \bibinfo {author}
  {\bibfnamefont {A.}~\bibnamefont {Banerjee}}, \bibinfo {author}
  {\bibfnamefont {P.}~\bibnamefont {Lampen-Kelley}}, \bibinfo {author}
  {\bibfnamefont {D.}~\bibnamefont {Mandrus}}, \bibinfo {author} {\bibfnamefont
  {S.~E.}\ \bibnamefont {Nagler}}, \bibinfo {author} {\bibfnamefont
  {S.}~\bibnamefont {Okamoto}}, \ and\ \bibinfo {author} {\bibfnamefont
  {D.~A.}\ \bibnamefont {Tennant}},\ }\href {\doibase
  10.1103/PhysRevResearch.4.L022061} {\bibfield  {journal} {\bibinfo  {journal}
  {Phys. Rev. Research}\ }\textbf {\bibinfo {volume} {4}},\ \bibinfo {pages}
  {L022061} (\bibinfo {year} {2022})}\BibitemShut {NoStop}%
\bibitem [{\citenamefont {Janssen}\ \emph {et~al.}(2017)\citenamefont
  {Janssen}, \citenamefont {Andrade},\ and\ \citenamefont
  {Vojta}}]{rucl3_janssen_17}%
  \BibitemOpen
  \bibfield  {author} {\bibinfo {author} {\bibfnamefont {L.}~\bibnamefont
  {Janssen}}, \bibinfo {author} {\bibfnamefont {E.~C.}\ \bibnamefont
  {Andrade}}, \ and\ \bibinfo {author} {\bibfnamefont {M.}~\bibnamefont
  {Vojta}},\ }\href {\doibase https://doi.org/10.1103/PhysRevB.96.064430}
  {\bibfield  {journal} {\bibinfo  {journal} {Phys. Rev. B}\ }\textbf {\bibinfo
  {volume} {96}},\ \bibinfo {pages} {064430} (\bibinfo {year}
  {2017})}\BibitemShut {NoStop}%
\bibitem [{\citenamefont {{Katukuri}}\ \emph {et~al.}(2015)\citenamefont
  {{Katukuri}}, \citenamefont {{Nishimoto}}, \citenamefont {{Rousochatzakis}},
  \citenamefont {{Stoll}}, \citenamefont {{van den Brink}},\ and\ \citenamefont
  {{Hozoi}}}]{rh213_vmk_15}%
  \BibitemOpen
  \bibfield  {author} {\bibinfo {author} {\bibfnamefont {V.~M.}\ \bibnamefont
  {{Katukuri}}}, \bibinfo {author} {\bibfnamefont {S.}~\bibnamefont
  {{Nishimoto}}}, \bibinfo {author} {\bibfnamefont {I.}~\bibnamefont
  {{Rousochatzakis}}}, \bibinfo {author} {\bibfnamefont {H.}~\bibnamefont
  {{Stoll}}}, \bibinfo {author} {\bibfnamefont {J.}~\bibnamefont {{van den
  Brink}}}, \ and\ \bibinfo {author} {\bibfnamefont {L.}~\bibnamefont
  {{Hozoi}}},\ }\href {\doibase 10.1038/srep14718} {\bibfield  {journal}
  {\bibinfo  {journal} {Sci. Rep.}\ }\textbf {\bibinfo {volume} {5}},\ \bibinfo
  {eid} {14718} (\bibinfo {year} {2015})}\BibitemShut {NoStop}%
\bibitem [{\citenamefont {{Bogdanov}}\ \emph {et~al.}(2015)\citenamefont
  {{Bogdanov}}, \citenamefont {{Katukuri}}, \citenamefont {{Romh{\'a}nyi}},
  \citenamefont {{Yushankhai}}, \citenamefont {{Kataev}}, \citenamefont
  {{B{\"u}chner}}, \citenamefont {{van den Brink}},\ and\ \citenamefont
  {{Hozoi}}}]{Ir214_niko_15}%
  \BibitemOpen
  \bibfield  {author} {\bibinfo {author} {\bibfnamefont {N.~A.}\ \bibnamefont
  {{Bogdanov}}}, \bibinfo {author} {\bibfnamefont {V.~M.}\ \bibnamefont
  {{Katukuri}}}, \bibinfo {author} {\bibfnamefont {J.}~\bibnamefont
  {{Romh{\'a}nyi}}}, \bibinfo {author} {\bibfnamefont {V.}~\bibnamefont
  {{Yushankhai}}}, \bibinfo {author} {\bibfnamefont {V.}~\bibnamefont
  {{Kataev}}}, \bibinfo {author} {\bibfnamefont {B.}~\bibnamefont
  {{B{\"u}chner}}}, \bibinfo {author} {\bibfnamefont {J.}~\bibnamefont {{van
  den Brink}}}, \ and\ \bibinfo {author} {\bibfnamefont {L.}~\bibnamefont
  {{Hozoi}}},\ }\href {\doibase 10.1038/ncomms8306} {\bibfield  {journal}
  {\bibinfo  {journal} {Nat. Commun.}\ }\textbf {\bibinfo {volume} {6}},\
  \bibinfo {pages} {7306} (\bibinfo {year} {2015})}\BibitemShut {NoStop}%
\bibitem [{\citenamefont {Wang}\ \emph {et~al.}(2021)\citenamefont {Wang},
  \citenamefont {Qi}, \citenamefont {Xi}, \citenamefont {Wang}, \citenamefont
  {Yu},\ and\ \citenamefont {Li}}]{PD}%
  \BibitemOpen
  \bibfield  {author} {\bibinfo {author} {\bibfnamefont {S.}~\bibnamefont
  {Wang}}, \bibinfo {author} {\bibfnamefont {Z.}~\bibnamefont {Qi}}, \bibinfo
  {author} {\bibfnamefont {B.}~\bibnamefont {Xi}}, \bibinfo {author}
  {\bibfnamefont {W.}~\bibnamefont {Wang}}, \bibinfo {author} {\bibfnamefont
  {S.-L.}\ \bibnamefont {Yu}}, \ and\ \bibinfo {author} {\bibfnamefont {J.-X.}\
  \bibnamefont {Li}},\ }\href {\doibase 10.1103/PhysRevB.103.054410} {\bibfield
   {journal} {\bibinfo  {journal} {Phys. Rev. B}\ }\textbf {\bibinfo {volume}
  {103}},\ \bibinfo {pages} {054410} (\bibinfo {year} {2021})}\BibitemShut
  {NoStop}%
\bibitem [{\citenamefont {Motrunich}(2005)}]{ring_exchange_motrunich_05}%
  \BibitemOpen
  \bibfield  {author} {\bibinfo {author} {\bibfnamefont {O.~I.}\ \bibnamefont
  {Motrunich}},\ }\href {\doibase https://doi.org/10.1103/PhysRevB.72.045105}
  {\bibfield  {journal} {\bibinfo  {journal} {Phys. Rev. B}\ }\textbf {\bibinfo
  {volume} {72}},\ \bibinfo {pages} {045105} (\bibinfo {year}
  {2005})}\BibitemShut {NoStop}%
\bibitem [{\citenamefont {Yadav}\ \emph {et~al.}(2019)\citenamefont {Yadav},
  \citenamefont {Eldeeb}, \citenamefont {Ray}, \citenamefont {Aswartham},
  \citenamefont {Sturza}, \citenamefont {Nishimoto}, \citenamefont {van~den
  Brink},\ and\ \citenamefont {Hozoi}}]{Yadav_CS}%
  \BibitemOpen
  \bibfield  {author} {\bibinfo {author} {\bibfnamefont {R.}~\bibnamefont
  {Yadav}}, \bibinfo {author} {\bibfnamefont {M.~S.}\ \bibnamefont {Eldeeb}},
  \bibinfo {author} {\bibfnamefont {R.}~\bibnamefont {Ray}}, \bibinfo {author}
  {\bibfnamefont {S.}~\bibnamefont {Aswartham}}, \bibinfo {author}
  {\bibfnamefont {M.~I.}\ \bibnamefont {Sturza}}, \bibinfo {author}
  {\bibfnamefont {S.}~\bibnamefont {Nishimoto}}, \bibinfo {author}
  {\bibfnamefont {J.}~\bibnamefont {van~den Brink}}, \ and\ \bibinfo {author}
  {\bibfnamefont {L.}~\bibnamefont {Hozoi}},\ }\href {\doibase
  10.1039/C8SC03018A} {\bibfield  {journal} {\bibinfo  {journal} {Chem. Sci.}\
  }\textbf {\bibinfo {volume} {10}},\ \bibinfo {pages} {1866} (\bibinfo {year}
  {2019})}\BibitemShut {NoStop}%
\bibitem [{\citenamefont {Winter}\ \emph {et~al.}(2016)\citenamefont {Winter},
  \citenamefont {Li}, \citenamefont {Jeschke},\ and\ \citenamefont
  {Valenti}}]{J3_winter_16}%
  \BibitemOpen
  \bibfield  {author} {\bibinfo {author} {\bibfnamefont {S.~M.}\ \bibnamefont
  {Winter}}, \bibinfo {author} {\bibfnamefont {Y.}~\bibnamefont {Li}}, \bibinfo
  {author} {\bibfnamefont {H.~O.}\ \bibnamefont {Jeschke}}, \ and\ \bibinfo
  {author} {\bibfnamefont {R.}~\bibnamefont {Valenti}},\ }\href {\doibase
  https://doi.org/10.1103/PhysRevB.93.214431} {\bibfield  {journal} {\bibinfo
  {journal} {Phys. Rev. B}\ }\textbf {\bibinfo {volume} {93}},\ \bibinfo
  {pages} {214431} (\bibinfo {year} {2016})}\BibitemShut {NoStop}%
\bibitem [{\citenamefont {White}(1992)}]{DMRG-1}%
  \BibitemOpen
  \bibfield  {author} {\bibinfo {author} {\bibfnamefont {S.~R.}\ \bibnamefont
  {White}},\ }\href {\doibase 10.1103/PhysRevLett.69.2863} {\bibfield
  {journal} {\bibinfo  {journal} {Phys. Rev. Lett.}\ }\textbf {\bibinfo
  {volume} {69}},\ \bibinfo {pages} {2863} (\bibinfo {year}
  {1992})}\BibitemShut {NoStop}%
\bibitem [{\citenamefont {Fishman}\ \emph {et~al.}(2022)\citenamefont
  {Fishman}, \citenamefont {White},\ and\ \citenamefont
  {Stoudenmire}}]{DMRG-2}%
  \BibitemOpen
  \bibfield  {author} {\bibinfo {author} {\bibfnamefont {M.}~\bibnamefont
  {Fishman}}, \bibinfo {author} {\bibfnamefont {S.~R.}\ \bibnamefont {White}},
  \ and\ \bibinfo {author} {\bibfnamefont {E.~M.}\ \bibnamefont
  {Stoudenmire}},\ }\href {\doibase 10.21468/SciPostPhysCodeb.4} {\bibfield
  {journal} {\bibinfo  {journal} {SciPost Phys. Codebases}\ ,\ \bibinfo {pages}
  {4}} (\bibinfo {year} {2022})}\BibitemShut {NoStop}%
\bibitem [{\citenamefont {Bogdanov}\ \emph {et~al.}(2013)\citenamefont
  {Bogdanov}, \citenamefont {Maurice}, \citenamefont {Rousochatzakis},
  \citenamefont {van~den Brink},\ and\ \citenamefont {Hozoi}}]{Os227_niko_12}%
  \BibitemOpen
  \bibfield  {author} {\bibinfo {author} {\bibfnamefont {N.~A.}\ \bibnamefont
  {Bogdanov}}, \bibinfo {author} {\bibfnamefont {R.}~\bibnamefont {Maurice}},
  \bibinfo {author} {\bibfnamefont {I.}~\bibnamefont {Rousochatzakis}},
  \bibinfo {author} {\bibfnamefont {J.}~\bibnamefont {van~den Brink}}, \ and\
  \bibinfo {author} {\bibfnamefont {L.}~\bibnamefont {Hozoi}},\ }\href@noop {}
  {\bibfield  {journal} {\bibinfo  {journal} {Phys. Rev. Lett.}\ }\textbf
  {\bibinfo {volume} {110}},\ \bibinfo {pages} {127206} (\bibinfo {year}
  {2013})}\BibitemShut {NoStop}%
\bibitem [{\citenamefont {Zangeneh}\ \emph {et~al.}(2019)\citenamefont
  {Zangeneh}, \citenamefont {Avdoshenko}, \citenamefont {van~den Brink},\ and\
  \citenamefont {Hozoi}}]{yb13_ziba_2019}%
  \BibitemOpen
  \bibfield  {author} {\bibinfo {author} {\bibfnamefont {Z.}~\bibnamefont
  {Zangeneh}}, \bibinfo {author} {\bibfnamefont {S.}~\bibnamefont
  {Avdoshenko}}, \bibinfo {author} {\bibfnamefont {J.}~\bibnamefont {van~den
  Brink}}, \ and\ \bibinfo {author} {\bibfnamefont {L.}~\bibnamefont {Hozoi}},\
  }\href {\doibase 10.1103/PhysRevB.100.174436} {\bibfield  {journal} {\bibinfo
   {journal} {Phys. Rev. B}\ }\textbf {\bibinfo {volume} {100}},\ \bibinfo
  {pages} {174436} (\bibinfo {year} {2019})}\BibitemShut {NoStop}%
\bibitem [{\citenamefont {Werner}\ \emph {et~al.}(2012)\citenamefont {Werner},
  \citenamefont {Knowles}, \citenamefont {Knizia}, \citenamefont {Manby},\ and\
  \citenamefont {Sch{\"u}tz}}]{Molpro}%
  \BibitemOpen
  \bibfield  {author} {\bibinfo {author} {\bibfnamefont {H.-J.}\ \bibnamefont
  {Werner}}, \bibinfo {author} {\bibfnamefont {P.~J.}\ \bibnamefont {Knowles}},
  \bibinfo {author} {\bibfnamefont {G.}~\bibnamefont {Knizia}}, \bibinfo
  {author} {\bibfnamefont {F.~R.}\ \bibnamefont {Manby}}, \ and\ \bibinfo
  {author} {\bibfnamefont {M.}~\bibnamefont {Sch{\"u}tz}},\ }\href {\doibase
  https://doi.org/10.1002/wcms.82} {\bibfield  {journal} {\bibinfo  {journal}
  {WIREs Comput. Mol. Sci.}\ }\textbf {\bibinfo {volume} {2}},\ \bibinfo
  {pages} {242} (\bibinfo {year} {2012})}\BibitemShut {NoStop}%
\bibitem [{\citenamefont {Klintenberg}\ \emph {et~al.}(2000)\citenamefont
  {Klintenberg}, \citenamefont {Derenzo},\ and\ \citenamefont
  {Weber}}]{Klintenberg_et_al}%
  \BibitemOpen
  \bibfield  {author} {\bibinfo {author} {\bibfnamefont {M.}~\bibnamefont
  {Klintenberg}}, \bibinfo {author} {\bibfnamefont {S.}~\bibnamefont
  {Derenzo}}, \ and\ \bibinfo {author} {\bibfnamefont {M.}~\bibnamefont
  {Weber}},\ }\href {\doibase https://doi.org/10.1016/S0010-4655(00)00071-0}
  {\bibfield  {journal} {\bibinfo  {journal} {Comp. Phys. Commun.}\ }\textbf
  {\bibinfo {volume} {131}},\ \bibinfo {pages} {120} (\bibinfo {year}
  {2000})}\BibitemShut {NoStop}%
\bibitem [{\citenamefont {Derenzo}\ \emph {et~al.}(2000)\citenamefont
  {Derenzo}, \citenamefont {Klintenberg},\ and\ \citenamefont
  {Weber}}]{Derenzo_et_al}%
  \BibitemOpen
  \bibfield  {author} {\bibinfo {author} {\bibfnamefont {S.~E.}\ \bibnamefont
  {Derenzo}}, \bibinfo {author} {\bibfnamefont {M.~K.}\ \bibnamefont
  {Klintenberg}}, \ and\ \bibinfo {author} {\bibfnamefont {M.~J.}\ \bibnamefont
  {Weber}},\ }\href {\doibase 10.1063/1.480776} {\bibfield  {journal} {\bibinfo
   {journal} {J. Chem. Phys.}\ }\textbf {\bibinfo {volume} {112}},\ \bibinfo
  {pages} {2074} (\bibinfo {year} {2000})}\BibitemShut {NoStop}%
\bibitem [{\citenamefont {Pipek}\ and\ \citenamefont
  {Mezey}(1989)}]{PM_method}%
  \BibitemOpen
  \bibfield  {author} {\bibinfo {author} {\bibfnamefont {J.}~\bibnamefont
  {Pipek}}\ and\ \bibinfo {author} {\bibfnamefont {P.~G.}\ \bibnamefont
  {Mezey}},\ }\href {\doibase 10.1063/1.456588} {\bibfield  {journal} {\bibinfo
   {journal} {J. Chem. Phys.}\ }\textbf {\bibinfo {volume} {90}},\ \bibinfo
  {pages} {4916} (\bibinfo {year} {1989})}\BibitemShut {NoStop}%
\bibitem [{\citenamefont {Ivanic}(2003)}]{ormas_ivanic_03}%
  \BibitemOpen
  \bibfield  {author} {\bibinfo {author} {\bibfnamefont {J.}~\bibnamefont
  {Ivanic}},\ }\href@noop {} {\bibfield  {journal} {\bibinfo  {journal} {J.
  Chem. Phys.}\ }\textbf {\bibinfo {volume} {119}},\ \bibinfo {pages} {9364}
  (\bibinfo {year} {2003})}\BibitemShut {NoStop}%
\bibitem [{\citenamefont {Bogdanov}\ \emph {et~al.}(2015)\citenamefont
  {Bogdanov}, \citenamefont {Katukuri}, \citenamefont {Romh{\'a}nyi},
  \citenamefont {Yushankhai}, \citenamefont {Kataev}, \citenamefont
  {B{\"u}chner}, \citenamefont {van~den Brink},\ and\ \citenamefont
  {Hozoi}}]{Bogdanov_et_al}%
  \BibitemOpen
  \bibfield  {author} {\bibinfo {author} {\bibfnamefont {N.~A.}\ \bibnamefont
  {Bogdanov}}, \bibinfo {author} {\bibfnamefont {V.~M.}\ \bibnamefont
  {Katukuri}}, \bibinfo {author} {\bibfnamefont {J.}~\bibnamefont
  {Romh{\'a}nyi}}, \bibinfo {author} {\bibfnamefont {V.}~\bibnamefont
  {Yushankhai}}, \bibinfo {author} {\bibfnamefont {V.}~\bibnamefont {Kataev}},
  \bibinfo {author} {\bibfnamefont {B.}~\bibnamefont {B{\"u}chner}}, \bibinfo
  {author} {\bibfnamefont {J.}~\bibnamefont {van~den Brink}}, \ and\ \bibinfo
  {author} {\bibfnamefont {L.}~\bibnamefont {Hozoi}},\ }\href {\doibase
  10.1038/ncomms8306} {\bibfield  {journal} {\bibinfo  {journal} {Nat.
  Commun.}\ }\textbf {\bibinfo {volume} {6}},\ \bibinfo {pages} {7306}
  (\bibinfo {year} {2015})}\BibitemShut {NoStop}%
\end{thebibliography}%


\begin{thebibliography}{21}%
\makeatletter
\providecommand \@ifxundefined [1]{%
 \@ifx{#1\undefined}
}%
\providecommand \@ifnum [1]{%
 \ifnum #1\expandafter \@firstoftwo
 \else \expandafter \@secondoftwo
 \fi
}%
\providecommand \@ifx [1]{%
 \ifx #1\expandafter \@firstoftwo
 \else \expandafter \@secondoftwo
 \fi
}%
\providecommand \natexlab [1]{#1}%
\providecommand \enquote  [1]{``#1''}%
\providecommand \bibnamefont  [1]{#1}%
\providecommand \bibfnamefont [1]{#1}%
\providecommand \citenamefont [1]{#1}%
\providecommand \href@noop [0]{\@secondoftwo}%
\providecommand \href [0]{\begingroup \@sanitize@url \@href}%
\providecommand \@href[1]{\@@startlink{#1}\@@href}%
\providecommand \@@href[1]{\endgroup#1\@@endlink}%
\providecommand \@sanitize@url [0]{\catcode `\\12\catcode `\$12\catcode
  `\&12\catcode `\#12\catcode `\^12\catcode `\_12\catcode `\%12\relax}%
\providecommand \@@startlink[1]{}%
\providecommand \@@endlink[0]{}%
\providecommand \url  [0]{\begingroup\@sanitize@url \@url }%
\providecommand \@url [1]{\endgroup\@href {#1}{\urlprefix }}%
\providecommand \urlprefix  [0]{URL }%
\providecommand \Eprint [0]{\href }%
\providecommand \doibase [0]{http://dx.doi.org/}%
\providecommand \selectlanguage [0]{\@gobble}%
\providecommand \bibinfo  [0]{\@secondoftwo}%
\providecommand \bibfield  [0]{\@secondoftwo}%
\providecommand \translation [1]{[#1]}%
\providecommand \BibitemOpen [0]{}%
\providecommand \bibitemStop [0]{}%
\providecommand \bibitemNoStop [0]{.\EOS\space}%
\providecommand \EOS [0]{\spacefactor3000\relax}%
\providecommand \BibitemShut  [1]{\csname bibitem#1\endcsname}%
\let\auto@bib@innerbib\@empty
%</preamble>
\bibitem [{\citenamefont {Werner}\ \emph {et~al.}(2012)\citenamefont {Werner},
  \citenamefont {Knowles}, \citenamefont {Knizia}, \citenamefont {Manby},\ and\
  \citenamefont {Sch{\"u}tz}}]{Molpro}%
  \BibitemOpen
  \bibfield  {author} {\bibinfo {author} {\bibfnamefont {H.-J.}\ \bibnamefont
  {Werner}}, \bibinfo {author} {\bibfnamefont {P.~J.}\ \bibnamefont {Knowles}},
  \bibinfo {author} {\bibfnamefont {G.}~\bibnamefont {Knizia}}, \bibinfo
  {author} {\bibfnamefont {F.~R.}\ \bibnamefont {Manby}}, \ and\ \bibinfo
  {author} {\bibfnamefont {M.}~\bibnamefont {Sch{\"u}tz}},\ }\href {\doibase
  https://doi.org/10.1002/wcms.82} {\bibfield  {journal} {\bibinfo  {journal}
  {WIREs Comput. Mol. Sci.}\ }\textbf {\bibinfo {volume} {2}},\ \bibinfo
  {pages} {242} (\bibinfo {year} {2012})}\BibitemShut {NoStop}%
\bibitem [{\citenamefont {Ortiz}\ \emph {et~al.}(shed)\citenamefont {Ortiz},
  \citenamefont {Sarte}, \citenamefont {Avidor}, \citenamefont {Hay},
  \citenamefont {Kenney}, \citenamefont {Kolisneikov}, \citenamefont
  {Pajerowski}, \citenamefont {Aczel}, \citenamefont {Taddei}, \citenamefont
  {Brown}, \citenamefont {Wang}, \citenamefont {Graf}, \citenamefont
  {Seshadri}, \citenamefont {Balents},\ and\ \citenamefont
  {Wilson}}]{ru112_Ortiz_2022}%
  \BibitemOpen
  \bibfield  {author} {\bibinfo {author} {\bibfnamefont {B.~R.}\ \bibnamefont
  {Ortiz}}, \bibinfo {author} {\bibfnamefont {P.~M.}\ \bibnamefont {Sarte}},
  \bibinfo {author} {\bibfnamefont {A.~H.}\ \bibnamefont {Avidor}}, \bibinfo
  {author} {\bibfnamefont {A.}~\bibnamefont {Hay}}, \bibinfo {author}
  {\bibfnamefont {E.}~\bibnamefont {Kenney}}, \bibinfo {author} {\bibfnamefont
  {A.~I.}\ \bibnamefont {Kolisneikov}}, \bibinfo {author} {\bibfnamefont
  {D.~M.}\ \bibnamefont {Pajerowski}}, \bibinfo {author} {\bibfnamefont
  {A.~A.}\ \bibnamefont {Aczel}}, \bibinfo {author} {\bibfnamefont {K.~M.}\
  \bibnamefont {Taddei}}, \bibinfo {author} {\bibfnamefont {C.}~\bibnamefont
  {Brown}}, \bibinfo {author} {\bibfnamefont {C.}~\bibnamefont {Wang}},
  \bibinfo {author} {\bibfnamefont {M.~J.}\ \bibnamefont {Graf}}, \bibinfo
  {author} {\bibfnamefont {R.}~\bibnamefont {Seshadri}}, \bibinfo {author}
  {\bibfnamefont {L.}~\bibnamefont {Balents}}, \ and\ \bibinfo {author}
  {\bibfnamefont {S.~D.}\ \bibnamefont {Wilson}},\ }\href
  {https://www.researchsquare.com/article/rs-1551865/v1} {\bibfield  {journal}
  {\bibinfo  {journal} {Research Square rs-1551865}\ } (\bibinfo {year}
  {unpublished})}\BibitemShut {NoStop}%
\bibitem [{\citenamefont {Klintenberg}\ \emph {et~al.}(2000)\citenamefont
  {Klintenberg}, \citenamefont {Derenzo},\ and\ \citenamefont
  {Weber}}]{Klintenberg_et_al}%
  \BibitemOpen
  \bibfield  {author} {\bibinfo {author} {\bibfnamefont {M.}~\bibnamefont
  {Klintenberg}}, \bibinfo {author} {\bibfnamefont {S.}~\bibnamefont
  {Derenzo}}, \ and\ \bibinfo {author} {\bibfnamefont {M.}~\bibnamefont
  {Weber}},\ }\href {\doibase https://doi.org/10.1016/S0010-4655(00)00071-0}
  {\bibfield  {journal} {\bibinfo  {journal} {Comp. Phys. Commun.}\ }\textbf
  {\bibinfo {volume} {131}},\ \bibinfo {pages} {120} (\bibinfo {year}
  {2000})}\BibitemShut {NoStop}%
\bibitem [{\citenamefont {Derenzo}\ \emph {et~al.}(2000)\citenamefont
  {Derenzo}, \citenamefont {Klintenberg},\ and\ \citenamefont
  {Weber}}]{Derenzo_et_al}%
  \BibitemOpen
  \bibfield  {author} {\bibinfo {author} {\bibfnamefont {S.~E.}\ \bibnamefont
  {Derenzo}}, \bibinfo {author} {\bibfnamefont {M.~K.}\ \bibnamefont
  {Klintenberg}}, \ and\ \bibinfo {author} {\bibfnamefont {M.~J.}\ \bibnamefont
  {Weber}},\ }\href {\doibase 10.1063/1.480776} {\bibfield  {journal} {\bibinfo
   {journal} {J. Chem. Phys.}\ }\textbf {\bibinfo {volume} {112}},\ \bibinfo
  {pages} {2074} (\bibinfo {year} {2000})}\BibitemShut {NoStop}%
\bibitem [{\citenamefont {Helgaker}\ \emph {et~al.}(2000)\citenamefont
  {Helgaker}, \citenamefont {J{\o}rgensen},\ and\ \citenamefont
  {Olsen}}]{olsen_bible}%
  \BibitemOpen
  \bibfield  {author} {\bibinfo {author} {\bibfnamefont {T.}~\bibnamefont
  {Helgaker}}, \bibinfo {author} {\bibfnamefont {P.}~\bibnamefont
  {J{\o}rgensen}}, \ and\ \bibinfo {author} {\bibfnamefont {J.}~\bibnamefont
  {Olsen}},\ }\href@noop {} {\emph {\bibinfo {title} {Molecular Electronic
  Structure Theory}}}\ (\bibinfo  {publisher} {John Wiley \& Sons},\ \bibinfo
  {address} {Chichester},\ \bibinfo {year} {2000})\BibitemShut {NoStop}%
\bibitem [{\citenamefont {Kreplin}\ \emph {et~al.}(2020)\citenamefont
  {Kreplin}, \citenamefont {Knowles},\ and\ \citenamefont
  {Werner}}]{MCSCF_Molpro}%
  \BibitemOpen
  \bibfield  {author} {\bibinfo {author} {\bibfnamefont {D.~A.}\ \bibnamefont
  {Kreplin}}, \bibinfo {author} {\bibfnamefont {P.~J.}\ \bibnamefont
  {Knowles}}, \ and\ \bibinfo {author} {\bibfnamefont {H.-J.}\ \bibnamefont
  {Werner}},\ }\href {\doibase 10.1063/1.5142241} {\bibfield  {journal}
  {\bibinfo  {journal} {J. Chem. Phys.}\ }\textbf {\bibinfo {volume} {152}},\
  \bibinfo {pages} {074102} (\bibinfo {year} {2020})}\BibitemShut {NoStop}%
\bibitem [{\citenamefont {Knowles}\ and\ \citenamefont
  {Werner}(1992)}]{MRCI_Molpro}%
  \BibitemOpen
  \bibfield  {author} {\bibinfo {author} {\bibfnamefont {P.~J.}\ \bibnamefont
  {Knowles}}\ and\ \bibinfo {author} {\bibfnamefont {H.-J.}\ \bibnamefont
  {Werner}},\ }\href {\doibase 10.1007/BF01117405} {\bibfield  {journal}
  {\bibinfo  {journal} {Theor. Chim. Acta}\ }\textbf {\bibinfo {volume} {84}},\
  \bibinfo {pages} {95} (\bibinfo {year} {1992})}\BibitemShut {NoStop}%
\bibitem [{\citenamefont {Berning}\ \emph {et~al.}(2000)\citenamefont
  {Berning}, \citenamefont {Schweizer}, \citenamefont {Werner}, \citenamefont
  {Knowles},\ and\ \citenamefont {Palmieri}}]{SOC_Molpro}%
  \BibitemOpen
  \bibfield  {author} {\bibinfo {author} {\bibfnamefont {A.}~\bibnamefont
  {Berning}}, \bibinfo {author} {\bibfnamefont {M.}~\bibnamefont {Schweizer}},
  \bibinfo {author} {\bibfnamefont {H.-J.}\ \bibnamefont {Werner}}, \bibinfo
  {author} {\bibfnamefont {P.~J.}\ \bibnamefont {Knowles}}, \ and\ \bibinfo
  {author} {\bibfnamefont {P.}~\bibnamefont {Palmieri}},\ }\href {\doibase
  10.1080/00268970009483386} {\bibfield  {journal} {\bibinfo  {journal} {Mol.
  Phys.}\ }\textbf {\bibinfo {volume} {98}},\ \bibinfo {pages} {1823} (\bibinfo
  {year} {2000})}\BibitemShut {NoStop}%
\bibitem [{\citenamefont {Peterson}\ \emph {et~al.}(2007)\citenamefont
  {Peterson}, \citenamefont {Figgen}, \citenamefont {Dolg},\ and\ \citenamefont
  {Stoll}}]{4d_elements}%
  \BibitemOpen
  \bibfield  {author} {\bibinfo {author} {\bibfnamefont {K.~A.}\ \bibnamefont
  {Peterson}}, \bibinfo {author} {\bibfnamefont {D.}~\bibnamefont {Figgen}},
  \bibinfo {author} {\bibfnamefont {M.}~\bibnamefont {Dolg}}, \ and\ \bibinfo
  {author} {\bibfnamefont {H.}~\bibnamefont {Stoll}},\ }\href {\doibase
  10.1063/1.2647019} {\bibfield  {journal} {\bibinfo  {journal} {J. Chem.
  Phys.}\ }\textbf {\bibinfo {volume} {126}},\ \bibinfo {pages} {124101}
  (\bibinfo {year} {2007})}\BibitemShut {NoStop}%
\bibitem [{\citenamefont {Dunning}(1989)}]{O-ligands}%
  \BibitemOpen
  \bibfield  {author} {\bibinfo {author} {\bibfnamefont {T.~H.}\ \bibnamefont
  {Dunning}},\ }\href {\doibase 10.1063/1.456153} {\bibfield  {journal}
  {\bibinfo  {journal} {J. Chem. Phys.}\ }\textbf {\bibinfo {volume} {90}},\
  \bibinfo {pages} {1007} (\bibinfo {year} {1989})}\BibitemShut {NoStop}%
\bibitem [{\citenamefont {Andrae}\ \emph {et~al.}(1990)\citenamefont {Andrae},
  \citenamefont {Haeussermann}, \citenamefont {Dolg}, \citenamefont {Stoll},\
  and\ \citenamefont {Preuss}}]{Andrae1990}%
  \BibitemOpen
  \bibfield  {author} {\bibinfo {author} {\bibfnamefont {D.}~\bibnamefont
  {Andrae}}, \bibinfo {author} {\bibfnamefont {U.}~\bibnamefont
  {Haeussermann}}, \bibinfo {author} {\bibfnamefont {M.}~\bibnamefont {Dolg}},
  \bibinfo {author} {\bibfnamefont {H.}~\bibnamefont {Stoll}}, \ and\ \bibinfo
  {author} {\bibfnamefont {H.}~\bibnamefont {Preuss}},\ }\href {\doibase
  10.1007/BF01114537} {\bibfield  {journal} {\bibinfo  {journal} {Theor. Chim.
  Acta}\ }\textbf {\bibinfo {volume} {77}},\ \bibinfo {pages} {123} (\bibinfo
  {year} {1990})}\BibitemShut {NoStop}%
\bibitem [{\citenamefont {Martin}\ and\ \citenamefont
  {Sundermann}(2001)}]{Martin_et_al}%
  \BibitemOpen
  \bibfield  {author} {\bibinfo {author} {\bibfnamefont {J.~M.~L.}\
  \bibnamefont {Martin}}\ and\ \bibinfo {author} {\bibfnamefont
  {A.}~\bibnamefont {Sundermann}},\ }\href {\doibase 10.1063/1.1337864}
  {\bibfield  {journal} {\bibinfo  {journal} {J. Chem. Phys.}\ }\textbf
  {\bibinfo {volume} {114}},\ \bibinfo {pages} {3408} (\bibinfo {year}
  {2001})}\BibitemShut {NoStop}%
\bibitem [{\citenamefont {Fuentealba}\ \emph {et~al.}(1982)\citenamefont
  {Fuentealba}, \citenamefont {Preuss}, \citenamefont {Stoll},\ and\
  \citenamefont {{Von Szentp{\'a}ly}}}]{Fuentealba_Na}%
  \BibitemOpen
  \bibfield  {author} {\bibinfo {author} {\bibfnamefont {P.}~\bibnamefont
  {Fuentealba}}, \bibinfo {author} {\bibfnamefont {H.}~\bibnamefont {Preuss}},
  \bibinfo {author} {\bibfnamefont {H.}~\bibnamefont {Stoll}}, \ and\ \bibinfo
  {author} {\bibfnamefont {L.}~\bibnamefont {{Von Szentp{\'a}ly}}},\ }\href
  {\doibase https://doi.org/10.1016/0009-2614(82)80012-2} {\bibfield  {journal}
  {\bibinfo  {journal} {Chem. Phys. Lett.}\ }\textbf {\bibinfo {volume} {89}},\
  \bibinfo {pages} {418} (\bibinfo {year} {1982})}\BibitemShut {NoStop}%
\bibitem [{\citenamefont {Pierloot}\ \emph {et~al.}(1995)\citenamefont
  {Pierloot}, \citenamefont {Dumez}, \citenamefont {Widmark},\ and\
  \citenamefont {Roos}}]{Pierloot1995}%
  \BibitemOpen
  \bibfield  {author} {\bibinfo {author} {\bibfnamefont {K.}~\bibnamefont
  {Pierloot}}, \bibinfo {author} {\bibfnamefont {B.}~\bibnamefont {Dumez}},
  \bibinfo {author} {\bibfnamefont {P.-O.}\ \bibnamefont {Widmark}}, \ and\
  \bibinfo {author} {\bibfnamefont {B.~O.}\ \bibnamefont {Roos}},\ }\href
  {\doibase 10.1007/BF01113842} {\bibfield  {journal} {\bibinfo  {journal}
  {Theor. Chim. Acta}\ }\textbf {\bibinfo {volume} {90}},\ \bibinfo {pages}
  {87} (\bibinfo {year} {1995})}\BibitemShut {NoStop}%
\bibitem [{\citenamefont {Pipek}\ and\ \citenamefont
  {Mezey}(1989)}]{PM_method}%
  \BibitemOpen
  \bibfield  {author} {\bibinfo {author} {\bibfnamefont {J.}~\bibnamefont
  {Pipek}}\ and\ \bibinfo {author} {\bibfnamefont {P.~G.}\ \bibnamefont
  {Mezey}},\ }\href {\doibase 10.1063/1.456588} {\bibfield  {journal} {\bibinfo
   {journal} {J. Chem. Phys.}\ }\textbf {\bibinfo {volume} {90}},\ \bibinfo
  {pages} {4916} (\bibinfo {year} {1989})}\BibitemShut {NoStop}%
\bibitem [{\citenamefont {Ivanic}(2003)}]{ormas_ivanic_03}%
  \BibitemOpen
  \bibfield  {author} {\bibinfo {author} {\bibfnamefont {J.}~\bibnamefont
  {Ivanic}},\ }\href@noop {} {\bibfield  {journal} {\bibinfo  {journal} {J.
  Chem. Phys.}\ }\textbf {\bibinfo {volume} {119}},\ \bibinfo {pages} {9364}
  (\bibinfo {year} {2003})}\BibitemShut {NoStop}%
\bibitem [{\citenamefont {Bogdanov}\ \emph {et~al.}(2015)\citenamefont
  {Bogdanov}, \citenamefont {Katukuri}, \citenamefont {Romh{\'a}nyi},
  \citenamefont {Yushankhai}, \citenamefont {Kataev}, \citenamefont
  {B{\"u}chner}, \citenamefont {van~den Brink},\ and\ \citenamefont
  {Hozoi}}]{Bogdanov_et_al}%
  \BibitemOpen
  \bibfield  {author} {\bibinfo {author} {\bibfnamefont {N.~A.}\ \bibnamefont
  {Bogdanov}}, \bibinfo {author} {\bibfnamefont {V.~M.}\ \bibnamefont
  {Katukuri}}, \bibinfo {author} {\bibfnamefont {J.}~\bibnamefont
  {Romh{\'a}nyi}}, \bibinfo {author} {\bibfnamefont {V.}~\bibnamefont
  {Yushankhai}}, \bibinfo {author} {\bibfnamefont {V.}~\bibnamefont {Kataev}},
  \bibinfo {author} {\bibfnamefont {B.}~\bibnamefont {B{\"u}chner}}, \bibinfo
  {author} {\bibfnamefont {J.}~\bibnamefont {van~den Brink}}, \ and\ \bibinfo
  {author} {\bibfnamefont {L.}~\bibnamefont {Hozoi}},\ }\href {\doibase
  10.1038/ncomms8306} {\bibfield  {journal} {\bibinfo  {journal} {Nat.
  Commun.}\ }\textbf {\bibinfo {volume} {6}},\ \bibinfo {pages} {7306}
  (\bibinfo {year} {2015})}\BibitemShut {NoStop}%
\bibitem [{\citenamefont {Yadav}\ \emph {et~al.}(2016)\citenamefont {Yadav},
  \citenamefont {Bogdanov}, \citenamefont {Katukuri}, \citenamefont
  {Nishimoto}, \citenamefont {van~den Brink},\ and\ \citenamefont
  {Hozoi}}]{Yadav2016}%
  \BibitemOpen
  \bibfield  {author} {\bibinfo {author} {\bibfnamefont {R.}~\bibnamefont
  {Yadav}}, \bibinfo {author} {\bibfnamefont {N.~A.}\ \bibnamefont {Bogdanov}},
  \bibinfo {author} {\bibfnamefont {V.~M.}\ \bibnamefont {Katukuri}}, \bibinfo
  {author} {\bibfnamefont {S.}~\bibnamefont {Nishimoto}}, \bibinfo {author}
  {\bibfnamefont {J.}~\bibnamefont {van~den Brink}}, \ and\ \bibinfo {author}
  {\bibfnamefont {L.}~\bibnamefont {Hozoi}},\ }\href {\doibase
  10.1038/srep37925} {\bibfield  {journal} {\bibinfo  {journal} {Sci. Rep.}\
  }\textbf {\bibinfo {volume} {6}},\ \bibinfo {pages} {37925} (\bibinfo {year}
  {2016})}\BibitemShut {NoStop}%
\bibitem [{\citenamefont {{Katukuri}}\ \emph {et~al.}(2015)\citenamefont
  {{Katukuri}}, \citenamefont {{Nishimoto}}, \citenamefont {{Rousochatzakis}},
  \citenamefont {{Stoll}}, \citenamefont {{van den Brink}},\ and\ \citenamefont
  {{Hozoi}}}]{rh213_vmk_15}%
  \BibitemOpen
  \bibfield  {author} {\bibinfo {author} {\bibfnamefont {V.~M.}\ \bibnamefont
  {{Katukuri}}}, \bibinfo {author} {\bibfnamefont {S.}~\bibnamefont
  {{Nishimoto}}}, \bibinfo {author} {\bibfnamefont {I.}~\bibnamefont
  {{Rousochatzakis}}}, \bibinfo {author} {\bibfnamefont {H.}~\bibnamefont
  {{Stoll}}}, \bibinfo {author} {\bibfnamefont {J.}~\bibnamefont {{van den
  Brink}}}, \ and\ \bibinfo {author} {\bibfnamefont {L.}~\bibnamefont
  {{Hozoi}}},\ }\href {\doibase 10.1038/srep14718} {\bibfield  {journal}
  {\bibinfo  {journal} {Sci. Rep.}\ }\textbf {\bibinfo {volume} {5}},\ \bibinfo
  {eid} {14718} (\bibinfo {year} {2015})}\BibitemShut {NoStop}%
\bibitem [{\citenamefont {Yadav}\ \emph {et~al.}(2019)\citenamefont {Yadav},
  \citenamefont {Eldeeb}, \citenamefont {Ray}, \citenamefont {Aswartham},
  \citenamefont {Sturza}, \citenamefont {Nishimoto}, \citenamefont {van~den
  Brink},\ and\ \citenamefont {Hozoi}}]{Yadav_CS}%
  \BibitemOpen
  \bibfield  {author} {\bibinfo {author} {\bibfnamefont {R.}~\bibnamefont
  {Yadav}}, \bibinfo {author} {\bibfnamefont {M.~S.}\ \bibnamefont {Eldeeb}},
  \bibinfo {author} {\bibfnamefont {R.}~\bibnamefont {Ray}}, \bibinfo {author}
  {\bibfnamefont {S.}~\bibnamefont {Aswartham}}, \bibinfo {author}
  {\bibfnamefont {M.~I.}\ \bibnamefont {Sturza}}, \bibinfo {author}
  {\bibfnamefont {S.}~\bibnamefont {Nishimoto}}, \bibinfo {author}
  {\bibfnamefont {J.}~\bibnamefont {van~den Brink}}, \ and\ \bibinfo {author}
  {\bibfnamefont {L.}~\bibnamefont {Hozoi}},\ }\href {\doibase
  10.1039/C8SC03018A} {\bibfield  {journal} {\bibinfo  {journal} {Chem. Sci.}\
  }\textbf {\bibinfo {volume} {10}},\ \bibinfo {pages} {1866} (\bibinfo {year}
  {2019})}\BibitemShut {NoStop}%
\bibitem [{\citenamefont {Janssen}\ \emph {et~al.}(2017)\citenamefont
  {Janssen}, \citenamefont {Andrade},\ and\ \citenamefont
  {Vojta}}]{rucl3_janssen_17}%
  \BibitemOpen
  \bibfield  {author} {\bibinfo {author} {\bibfnamefont {L.}~\bibnamefont
  {Janssen}}, \bibinfo {author} {\bibfnamefont {E.~C.}\ \bibnamefont
  {Andrade}}, \ and\ \bibinfo {author} {\bibfnamefont {M.}~\bibnamefont
  {Vojta}},\ }\href {\doibase https://doi.org/10.1103/PhysRevB.96.064430}
  {\bibfield  {journal} {\bibinfo  {journal} {Phys. Rev. B}\ }\textbf {\bibinfo
  {volume} {96}},\ \bibinfo {pages} {064430} (\bibinfo {year}
  {2017})}\BibitemShut {NoStop}%
\end{thebibliography}%

\end{document}

\title{Supplementary Information --- NaRuO$_2$: Kitaev-Heisenberg exchange in triangular-lattice setting}

\author{Pritam Bhattacharyya}
\affiliation{Institute for Theoretical Solid State Physics, Leibniz IFW Dresden, Helmholtzstra{\ss}e~20, 01069 Dresden, Germany}

\author{Nikolay Bogdanov}
\affiliation{Max Planck Institute for Solid State Research, Heisenbergstra{\ss}e~1, 70569 Stuttgart, Germany}

\author{Satoshi Nishimoto}
\affiliation{Institute for Theoretical Solid State Physics, Leibniz IFW Dresden, Helmholtzstra{\ss}e~20, 01069 Dresden, Germany}
\affiliation{Department of Physics, Technical University Dresden, 01069 Dresden, Germany}

\author{Stephen D.~Wilson}
\affiliation{Materials Department, University of California, Santa Barbara, California 93106-5050, USA}

\author{Liviu Hozoi}
\affiliation{Institute for Theoretical Solid State Physics, Leibniz IFW Dresden, Helmholtzstra{\ss}e~20, 01069 Dresden, Germany}

\date\today
\maketitle

{\it Quantum chemical computational details.\,}
All quantum chemical computations were carried out using the {\sc molpro} suite of programs \cite{
Molpro}.
Crystallographic data as reported by Ortiz {\it et al.}~\cite{ru112_Ortiz_2022} were utilized.
For each type of embedded cluster, the crystalline environment was modeled as a large array of point
charges which reproduces the crystalline Madelung field within the cluster volume; 
we employed the {\sc ewald} program \cite{Klintenberg_et_al,Derenzo_et_al} to generate the point-charge
embeddings.

\begin{figure*}[!bh!]
        \includegraphics[width=1.90\columnwidth]{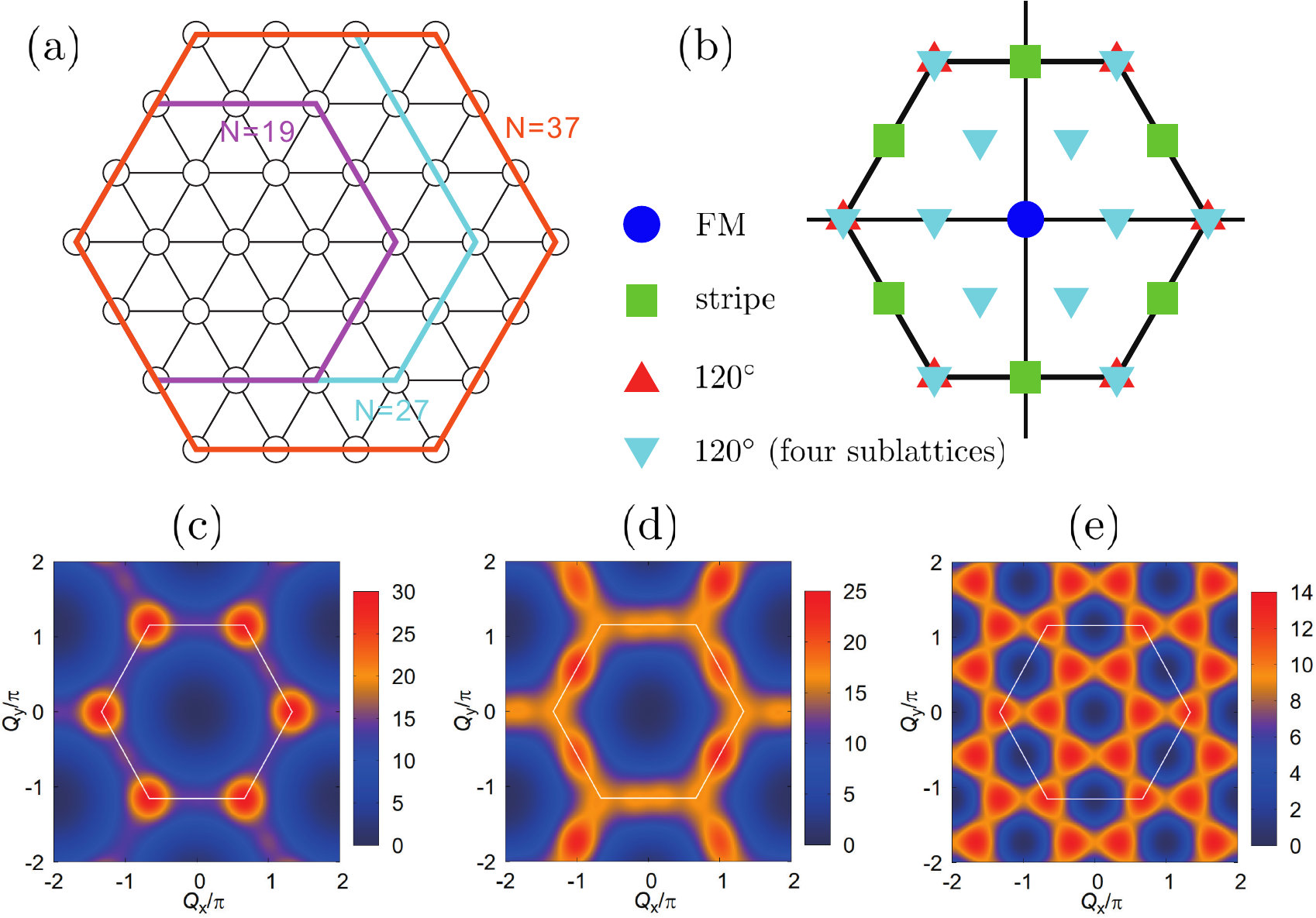}
        \caption{
(a) Finite fragments (with open boundary conditions) used in the 
DMRG computations.
(b) Brillouin zone of the triangular lattice with ordering vectors for a few relevant phases.
Static spin structure factors for 
(c) $J=1$, (d) $J=1$, $J_2=0.3$, and (e) $J=-1$, $J_3=10$ (meV).
        }
\label{fig_dmrg_sm}
\end{figure*}

\begin{figure*}[!bh!]
\includegraphics[width=1.90\columnwidth]{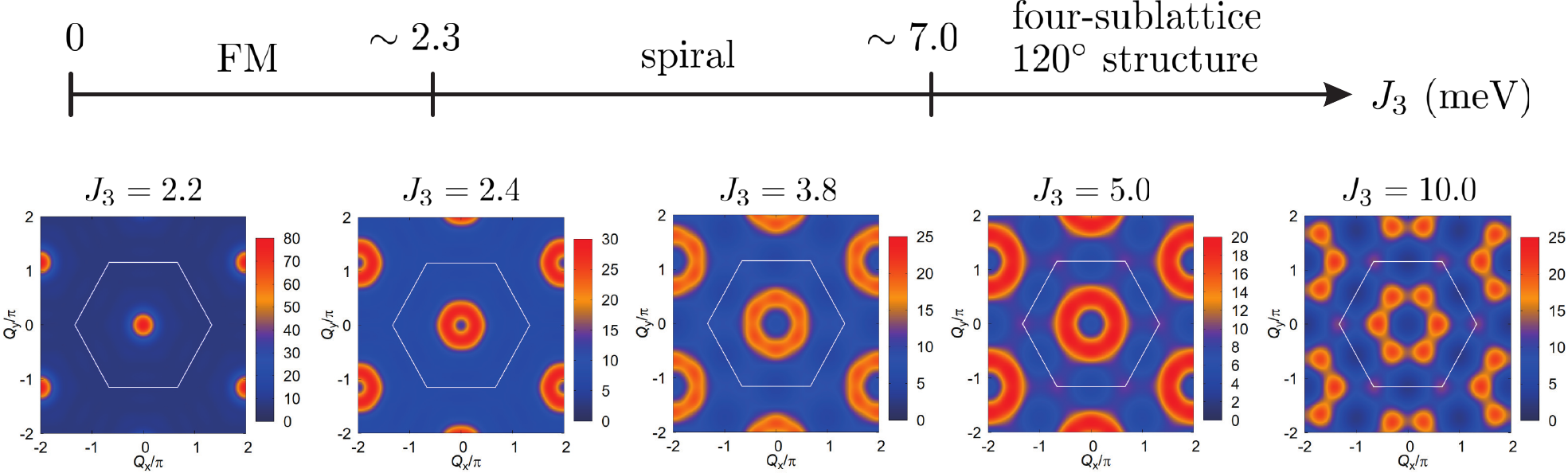}
\caption{
	Ground-state phase diagram obtained by DMRG computations for a $J$-$K$-$\Gamma$-$\Gamma'$-$J_3$ model
	(top) and static spin structure factor for variable third-neighbor Heisenberg coupling $J_3$ (bottom).
	Using the quantum chemical analysis as a guide, $J$, $K$,
	$\Gamma$, and $\Gamma'$ were set to to --5.5, 4, 4, and 1.5 meV, respectively.
	}
\label{fig_PDJ3_sm}
\end{figure*}

\begin{figure*}[!bh!]
	\includegraphics[width=1.90\columnwidth]{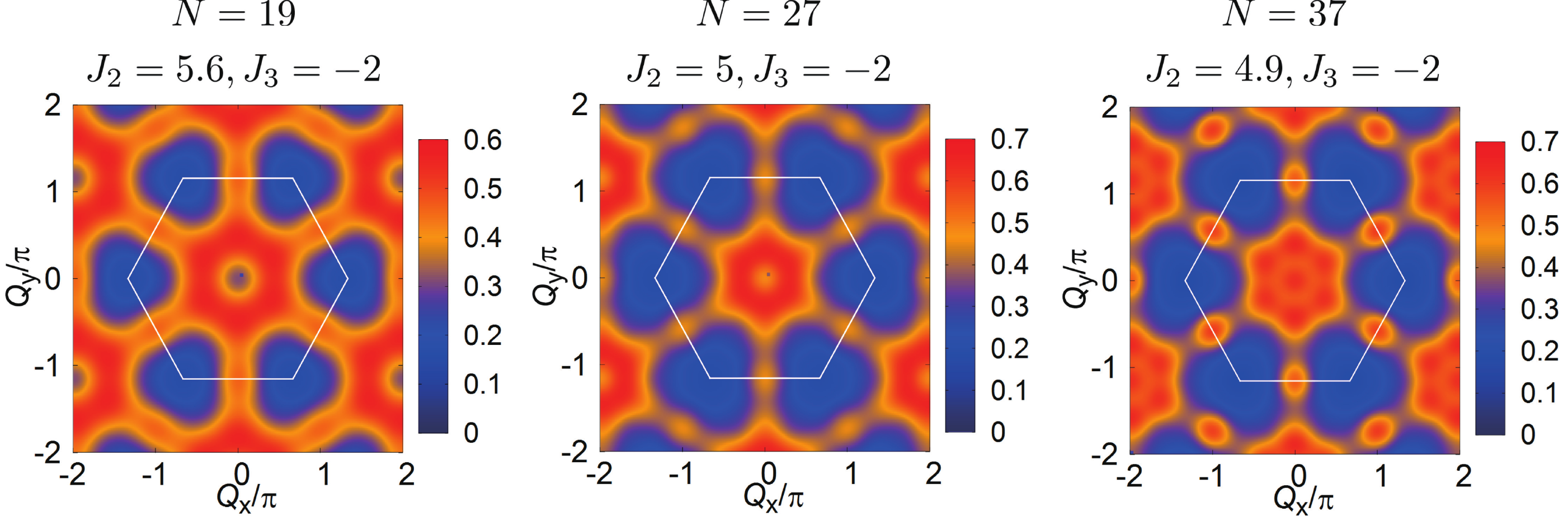}
	\caption{
Static spin structure factor characterizing the QSL state in the vicinity of the critical point within the AF-$J_2$, FM-$J_3$ quadrant. 
	}
	\label{fig_PPSQ_sm}
\end{figure*}

To clarify the Ru-site multiplet structure, a cluster consisting of one `central' RuO$_6$ octahedron,
the six in-plane adjacent octahedra, and 12 nearby Na cations was designed.
The quantum chemical study was initiated as complete active space self-consistent field (CASSCF)
calculations \cite{olsen_bible,MCSCF_Molpro}, with an active orbital space containing the five 4$d$
orbitals of the central Ru ion.
Post-CASSCF correlation computations were performed at the level of multireference configuration-interaction
(MRCI) with single and double excitations \cite{olsen_bible,MRCI_Molpro} out of the Ru 4$d$ and O 2$p$
orbitals of the central RuO$_6$ octahedron.
Spin-orbit couplings (SOCs) were accounted for following the procedure described in Ref.~\cite{SOC_Molpro}. 
For the central Ru ion energy-consistent relativistic pseudopotentials (ECP28MDF) and Gaussian-type
valence basis sets of effective quadruple-$\zeta$ quality (referred to as ECP28MDF-VTZ in the {\sc
molpro} library) \cite{4d_elements} were employed, whereas we used all-electron triple-$\zeta$ basis
sets for the O ligands of the central RuO$_6$ octahedron \cite{O-ligands}. 
The six adjacent in-plane cations were represented as closed-shell Rh$^{3+}$ $t_{2g}^6$ species, using
quasirelativistic pseudopotentials (Ru ECP29MWB) \cite{Andrae1990} and (Ru ECP28MWB) (8s7p6d)/[5s4p3d]
basis sets for electrons in the 4th shell \cite{Andrae1990,Martin_et_al}.
Large-core pseudopotentials were adopted for the 12 adjacent Na cations \cite{Fuentealba_Na}. 
 
Clusters with two edge-sharing RuO$_6$ octahedra in the central region were considered in order to
to derive the inter-site effective magnetic couplings.
The eight in-plane octahedra directly linked to the two-octahedra central unit and 22 nearby Na ions
were also explicitly included in the quantum chemical computations but using much more compact basis
sets.
We utilized energy-consistent relativistic pseudopotentials (ECP28MDF) and Gaussian-type valence basis
sets of effective quadruple-$\zeta$ quality (ECP28MDF-VTZ) for the central Ru species \cite{4d_elements}.
All-electron basis sets of quintuple-$\zeta$ quality were employed for the two bridging ligands
\cite{O-ligands} while all-electron basis sets of triple-$\zeta$ quality were applied for the remaining
eight O anions \cite{O-ligands} associated with the two octahedra of the reference unit.
The eight adjacent transition-metal (TM) cations were represented as closed-shell Rh$^{3+}$ $t_{2g}^6$
species, using non-relativistic pseudopotentials (Ru ECP29MHF) and (Ru ECP28MHF) (8s7p6d)/[4s3p2d] basis
sets \cite{Andrae1990,Martin_et_al};
the outer 22 O ligands associated with the eight adjacent octahedra were described through minimal
all-electron atomic-natural-orbital (ANO) basis sets \cite{Pierloot1995}. 
Large-core pseudopotentials were considered for the 22 Na nearby cations \cite{Fuentealba_Na}.

CASSCF computations were carried out with six (Ru $t_{2g}$) valence orbitals and ten electrons as
active (abbreviated as (10e,6o) active space);
the $t_{2g}$ orbitals of the adjacent TM ions were part of the inactive orbital space.
In the subsequent MRCI correlation treatment, single and double excitations out of the central-unit
Ru $t_{2g}$ and bridging-O 2$p$ levels were considered.
We used the Pipek-Mezey methodology \cite{PM_method} to obtain localized central-unit orbitals.

The CASSCF optimization was performed for the lowest nine singlet and lowest nine triplet states
associated with the (10e,6o) setting.
Those were the states for which SOCs were further accounted for \cite{SOC_Molpro}, either at 
single-configuration (SC), CASSCF, or MRCI level, which finally yields a number of 36 spin-orbit
states.
The SC label in Table III in the main text stands for a CASCI in which intersite excitations 
are not considered.
This is also referred to as occupation-restricted multiple active space (ORMAS) scheme \cite{
ormas_ivanic_03}.

Only one type of Ru-Ru links is present in NaRuO$_2$.
The unit of two nearest-neighbor octahedra displays $C_{2h}$ point-group symmetry, which implies a 
generalized bilinear effective spin Hamiltonian of the following form for a pair of adjacent
1/2-pseudospins ${\bf{\tilde{S}}}_i$ and ${\bf{\tilde{S}}}_j$\,:
\begin{equation}
\label{eqn:Hamil}
\mathcal{H}_{ij}^{(\gamma)} = J{\bf{\tilde{S}}}_i\cdot {\bf{\tilde{S}}}_j + \\
                              K\tilde{S}_i^{\gamma}\tilde{S}_j^{\gamma} + \\
                              \sum_{\alpha\neq\beta} \Gamma_{\alpha\beta} \\
                              (\tilde{S}_i^{\alpha}\tilde{S}_j^{\beta} +  \\
                              \tilde{S}_i^{\beta}\tilde{S}_j^{\alpha}).
\end{equation}
The $\Gamma_{\alpha\beta}$ coefficients refer to the off-diagonal components of the 3$\times$3
symmetric-anisotropy exchange matrix;
$\alpha,\beta,\gamma\!\in\!\{x,y,z\}$.
An antisymmetric Dzyaloshinskii-Moriya coupling is not allowed, given the inversion center.
% A local Kitaev reference frame is used here, such that for each Ru--Ru link the $z$ axis is
% perpendicular to the `magnetic' Ru$_2$O$_2$ plaquette.

The lowest four spin-orbit eigenstates from the {\sc molpro} output (eigenvalues lower by $\sim$200
meV or more than the eigenvalues of higher-lying excited states, as illustrated for example in Table\;I)
were mapped onto the eigenvectors of the effective spin Hamiltonian (\ref{eqn:Hamil}),
following the procedure described in Refs.~\cite{Bogdanov_et_al,Yadav2016}\,:
those four expectation values and the matrix elements of the Zeeman Hamiltonian in the basis of the
four lowest-energy spin-orbit eigenvectors are put in direct correspondence with the respective
eigenvalues and matrix elements of (\ref{eqn:Hamil}).
Having two of the states in the same irreducible representation of the $C_{2h}$ point group \cite{
rh213_vmk_15}, such one-to-one mapping translates into two possible sets of effective magnetic
couplings.
The relevant array is chosen as the one whose $g$ factors fit the $g$ factors obtained 
for a single RuO$_6$ $t_{2g}^5$ octahedron. 

We used the standard coordinate frame usually employed in the literature, different from the rotated
frame employed in earlier quantum chemical studies \cite{Yadav2016,rh213_vmk_15,Yadav_CS} that
affects the sign of $\Gamma$.
This is the reason the sign of $\Gamma$ for RuCl$_3$ in Table II in the main text is different from
the sign in Ref.~\cite{Yadav2016} (see also footnote [48] in  Ref.~\cite{rucl3_janssen_17}).

{\it Effective-model computations.\,}
To demonstrate that various magnetic phases in the triangular-lattice effective-spin model can be
properly identified for the clusters employed in our DMRG calculations (see Fig.\;\ref{fig_dmrg_sm}(a)\,),
we plot in Fig.\;\ref{fig_dmrg_sm}(c-e) the static spin structure factor
\begin{equation}
      S({\bf Q})=\frac{1}{N}\sum_{ij} \langle {\tilde {\bf S}}_i\cdot {\tilde {\bf S}}_j \rangle \exp[i {\bf Q}\cdot ({\bf r}_i-{\bf r}_j)]
\end{equation}
for a $120^\circ$ structure (with $J=1$), stripe order (computed for $J=1$, $J_2=0.3$), and a
four-sublattice $120^\circ$ structure ($J=-1$, $J_3=10$).
In the expression for $S({\bf Q})$, $N$ is the number of sites and ${\bf r}_i$ is the position of site
$i$.
The Brillouin zone and the relevant ordering vectors are depicted in Fig.\;\ref{fig_dmrg_sm}(b).
%
% {\it Propagation vector of the spiral state.\,}
As mentioned in the main text, in the spiral phase the propagation vector increases continuously when
moving away from the ferromagnetic (FM) sector.
To illustrate this, we show in Fig.\;\ref{fig_PDJ3_sm} the evolution of $S({\bf Q})$ for variable $J_3$
and $J_2\!=\!0$ ($J$, $K$, $\Gamma$, and $\Gamma'$ being set to --5.5, 4, 4, and 1.5 meV, respectively).
We find that the propagation vector increases continuously with increasing $J_3$, from $Q\!=\!0$
at the FM-spiral critical point $J_3\!\approx\!2.3$ to $Q\!=\!2\pi/3$, corresponding to the Bragg peak of
a four-sublattice 120$^{\circ}$ structure, at $J_3\!\approx\!7$.

{\it Cluster-size dependence of $S({\bf Q})$ for $J_2\!\approx\!5$, $J_3\!\approx\!-2$.\,}
For longer-range Heisenberg couplings $J_2\!\approx\!5$ and $J_3\!\approx\!-2$, inside the QSL phase,
the structure factor exhibits a broad spectral feature for momenta near $Q\!=\!0$, see discussion
in the main text.
Four ordered phases are contiguous at this spot, namely, the FM, spiral, stripy, and two-sublattice
120$^\circ$ phases.
Inside the QSL phase, close to this critical point, the overall appearance of $S({\bf Q})$ for
$Q\!\approx\!0$ does not depend on the size of the employed cluster, as illustrated in
Fig.\;\ref{fig_PPSQ_sm}.
The value of $J_2$ is slightly different for each system size due to the finite-size effects.
% The values of $J_2$ and $J_3$ for $N=19$ are slightly tuned due to the finite-size effects.

\bibliography{refs_apr19}